\documentclass[a4,12pt,epsf]{article}

\usepackage{graphicx,amsmath}
\usepackage{amsmath,amsfonts,latexsym,amssymb}
\usepackage{verbatim,amsthm,curves,graphics}
\usepackage{mathrsfs}

\def\b{s}
\def\ben{\begin{equation}}
\def\een{\end{equation}}
\def\bena{\begin{eqnarray}}
\def\eena{\end{eqnarray}}
\def\non{\nonumber}
\def\d{{\rm d}}
\def\N{{\cal N}}

\def\D{D}
\def\E{{\bf E}}
\def\mr{{\mathbb R}}
\def\A{{\bf A}}
\def\B{{\bf B}}
\def\Q{{\bf Q}}
\def\J{{\bf J}}
\def\C{{\bf C}}
\def\Y{{\bf Y}}

\newcommand{\LL}{{\bf L}}
\newcommand{\bt}{\mbox{\boldmath $\theta$}}
\newcommand{\btau}{\mbox{\boldmath $\tau$}}
\newcommand{\bomega}{\mbox{\boldmath $\omega$}}
\newcommand{\beps}{\mbox{\boldmath $\epsilon$}}
\newcommand{\I}{{\mathcal I}}
\renewcommand{\H}{{\mathcal H}}
\newcommand{\F}{{\mathcal F}}

\newcommand{\R}{{\mathcal R}}

\begin{document}
%---------------------------------------------------------------------------

\title{Asymptotic generators of fermionic charges and boundary
  conditions preserving supersymmetry}

\author{Stefan Hollands\thanks{hollands@theorie.physik.uni-goe.de}\\
Institut f\" ur Theoretische Physik\\
Georg-August Universit\" at, D-37077 G\" ottingen, Germany\\
\and
Donald Marolf\thanks{marolf@physics.ucsb.edu}\\
Department of Physics, UCSB, Santa Barbara, CA 93106
}

\maketitle

\begin{abstract}

We use a covariant phase space formalism to give a general prescription
for defining
Hamiltonian generators of bosonic and fermionic symmetries in
diffeomorphism invariant theories, such as supergravities.
A simple and general criterion is derived for a choice of boundary
condition to lead to conserved generators of the symmetries on the
phase space.  In particular, this provides a criterion for the
preservation of supersymmetries.
For bosonic symmetries corresponding to diffeomorphisms,
our prescription coincides with the
method of Wald et al.

We then illustrate these methods in the case of certain supergravity
theories in $d=4$. In minimal AdS supergravity, boundary conditions
such that the supercharges exist as Hamiltonian generators of
supersymmetry transformations are unique within the usual framework
in which the boundary metric is fixed. In extended ${\mathcal N}=4$
AdS supergravity, or more generally in the presence of chiral matter
superfields, we find that there exist many boundary conditions
preserving ${\mathcal N}=1$ supersymmetry for which corresponding
generators exist. These choices are shown to correspond to a choice
of certain arbitrary boundary ``superpotentials,'' for suitably
defined ``boundary superfields.'' We also derive corresponding
formulae for the conserved bosonic charges, such as energy, in those
theories, and we argue that energy is always positive, for any
supersymmetry-preserving boundary conditions. We finally comment on
the relevance and interpretation of our results within the AdS-CFT
correspondence.
\end{abstract}

\pagebreak

\section{Introduction}

In  a classical system
 defined by a Hamiltonian or Lagrangian, there
typically exists a constant of the motion for any symmetry of the
system. Furthermore, any such conserved quantity generates the
action of the symmetry on the phase space of the theory.
While this result is simple and straightforward to derive for
mechanical systems with a finite number of degrees of freedom, the
situation can be considerably more complicated in field theories. This
is in particular the case for field theories on manifolds with
asymptotic regions or boundaries, where the existence and form of the
conserved quantities will in general depend in a subtle way on the
precise choice of the phase space of the theory, i.e., the asymptotic
or boundary conditions that one chooses to impose upon the fields.

Of particular interest in theoretical physics are theories with local
gauge invariance, such as diffeomorphism invariance, Yang-Mills type
gauge invariance, or local supersymmetry. For theories with
diffeomorphism invariance,
% such as general relativity,
a general analysis of the existence of conserved quantities within a
``covariant phase space framework'' was given by Wald et
al.~\cite{Lee1989,Iyer1994,Iyer1995,Zoupas1998}, (see
also~\cite{Ashtekar1,Ashtekar2,Chrusciel}). In these works, a
general criterion was derived stating when a given boundary or
asymptotic condition on the fields allows the existence of conserved
charges generating asymptotic spacetime symmetries. These ideas were
applied e.g. in~\cite{Hollands2003} to derive formulae for the Bondi
energy in higher dimensional general relativity with asymptotically
flat asymptotic conditions, and in~\cite{marolf, KS9,
Hertog2005,Amsel2006} to derive expressions for  bosonic  conserved
charges (such as  energy) in gravity theories with asymptotically
AdS boundary conditions.

In this paper, we generalize the analysis of Wald et al. to
charges of fermionic type. Our first result is a general criterion  for
when a given boundary or asymptotic condition on the fields
permits the definition of conserved asymptotic charges conjugate to
 a fermionic asymptotic symmetry.
That condition is analogous to the one for bosonic symmetries and
can be simply stated as follows: (1) The symplectic flux through the
boundary (or at infinity) has to vanish under the imposed boundary
(or asymptotic) conditions. (2) The boundary (or asymptotic)
conditions must be invariant under the asymptotic symmetry.

We then
%to
apply this general formalism
to supergravity theories with a negative cosmological constant, for
which asymptotically AdS boundary conditions are possible. Our
first example is minimal  $\N=1$ supergravity~\cite{Townsend1977}.  Within the usual framework where the boundary metric is held fixed, we  show that there exists a unique boundary condition on the metric
and the spin-3/2 field so that conditions (1) and (2) are satisfied
for local supersymmetry transformations. We also derive the
expression for the corresponding supercharge, which is
found to agree with an expression derived earlier by a somewhat
different method~\cite{Henneaux1985}.

Next, we investigate extended $\N=4$
supergravity~\cite{Freedman1977} in a scalar
reduction~\cite{Das1977} in which, besides the gravity multiplet,
only the scalar multiplet is kept. The scalar fields have a mass
$m^2=-2$, which is within the Breitenlohner-Freedman
range~\cite{BF,IW}. The reduced theory has a local $\N=1$
invariance, and we investigate possible boundary conditions such
that there exist corresponding Hamiltonian generators for the $\N=1$
supersymmetry transformations. Here, we find that there is a
1-parameter family of boundary conditions on the fields in the
gravity and scalar multiplets such that (1) and (2) are satisfied.
Therefore, conserved generators exist for those boundary conditions.
The boundary conditions found by~\cite{BF} correspond to 2
particular extreme values of the parameter. We also investigate
boundary conditions preserving only a 2-dimensional subspace of the
fermionic $\N=1$ symmetry generating asymptotic
 Poincare transformations
in the ``Poincare
compactification'' of AdS-spacetime. Here, we find that there exist
 additional boundary conditions satisfying (1) and (2). These boundary
conditions can be formulated as a
 ``supersymmetric Lagrange submanifold condition,''
 in terms of an arbitrarily chosen ``boundary
superpotential'' of suitably defined ``boundary superfields,'' which
in turn are built from the asymptotic values of the fields in the scalar
multiplet (in a somewhat non-obvious way). These boundary conditions
preserve an $O(2,1) \times \mr^3$ of the standard $O(3,2)$ bosonic
symmetry group.

 Finally, we construct the explicit expressions for the bosonic and fermionic
conserved charges in the scalar reduction of the $\N=4$ extended
supergravity theory. We find that the expression for the bosonic
charges contains a piece that is a surface integral at infinity of the
electric Weyl tensor, as well as an extra piece involving the fields in the
scalar multiplet. In the case of $O(2,1) \times \mr^3$ invariant
boundary conditions, the extra piece involves the boundary
superpotential of the fields in the scalar multiplet. In the special
case of the boundary conditions found by Breitenlohner and
Freedman~\cite{BF}, only the Weyl piece remains.

Our results have the following standard interpretation in the context of the
AdS/CFT-correspondence~\cite{Maldacena:1997re, Gubser:1998bc,Witten:1998qj,Aharony:1999ti}. In the $\N=4$-theory,
%% the choice of
%% boundary conditions other than those described in~\cite{BF}
changing the boundary conditions corresponds to adding a term
to the action of the boundary quantum field theory. Since our
boundary conditions are formulated in terms of a supersymmetric
Lagrange submanifold condition, the additional piece is given
precisely~\cite{Witten:2001ua,Berkooz:2002ug,Sever:2002fk} by the
boundary superpotential of the boundary fields corresponding to the
scalar multiplet. Such deformations of AdS/CFT have been of
significant interest, see e.g., \cite{PS,PW, LM}.

\section{General definition of fermionic charges}
\label{defs}
 Assume we are given a Lagrange density $\LL$, (viewed
as a $d$-form) on a $d$-dimensional manifold
$M$. We assume that $\LL$ depends on
a metric $g_{\mu\nu}$,
tensor fields $\phi_{\mu_1 \dots \mu_k}$, gauge fields $A_\mu$, and
mixed spinor-tensor fields $\psi^{A_1 \dots A_q}{}_{\nu_1 \dots \nu_p}$.
We shall abbreviate the collection of all dynamical fields by
\ben
\Phi = (g_{\mu\nu}, \phi_{\mu_1 \mu_2 \dots \mu_k}, A_\mu,
\psi^{A_1 A_2 \dots A_q}{}_{\nu_1 \nu_2 \dots \nu_p}) \, .
\een
  To define spinors we assume that a spin structure exists and that one has been chosen
  as part of the definition of the
theory. In quantum field theory, by the spin-statistics theorem,
fields of half odd integer spin must be anticommuting, while fields
of integer spin must be commuting. But in classical field theories,
there is a choice to be made whether one wants to view the spinor
fields as commuting or anti-commuting, and there does not appear to
be a good physical justification for either choice. However, studies
of the initial value problem in supergravity
theory~\cite{Choquet1983,Isenberg1983} show that such a theory
possess a well-posed initial value formulation if spinors are taken
to anti-commute, while difficulties seem to arise if one assumes
them to be commuting. We shall therefore assume this in our paper,
too, i.e., half odd integer spin fields are taken as anti-commuting
and integer spin fields are taken to commute. Concretely, this may
e.g. be implemented~\cite{Bao1984} by taking the fields to be valued
in an exterior coefficient algebra $A={\rm Ext}(V)$, where $V$ is an
auxiliary infinite dimensional vector space, tensored with the
appropriate vector bundles corresponding to the vector or spinor
nature of the field. From this perspective, a field is then a formal
power series \ben \Phi(x) = \sum_{n=0}^\infty v_{(n)}\Phi_{(n)}(x)
\, , \een where $v_{(n)}$ is an $n$-th exterior power in $A$.
Furthermore, it is understood that a half odd integer field
component of $\Phi$ only contains terms with odd $n$ in the series,
while a field with integer spin only contains terms with even $n$.
As shown in~\cite{Choquet1983, Isenberg1983} for $\N=1$
supergravity, solutions to the field equations then exist as formal
power series in this algebra. The first term $\Phi_{(0)}$ is the
zeroth order metric, which is seen to satisfy the Einstein equation
with all spinors set to 0. Because $A=A_+ \oplus A_-$ may be
decomposed into even (bosonic) and odd (fermionic) elements, we have
a corresponding decomposition of field quantities.

We assume that $\LL$ is commuting and
does not depend on any non-dynamical background
fields, or equivalently, that $\LL$ is fully
diffeomorphism covariant, in the sense that\footnote{Note that
it does not make to speak about the pull back of a spinor field
separately. But it makes sense if we also pull back simultaneously
the metric and corresponding
spinor bundles with the diffeomorphism. This is what is meant in this
formula, and similar other formulae below.}
\ben
f^* \LL(\Phi) = \LL(f^* \Phi)
\een
for any diffeo $f$ on $M$.
On the space of (smooth) fields,
we will consider various operations. Given a functional $F$
on the space of fields and a smooth 1-parameter family $\Phi_t$
in the space of fields with $\Phi_0 = \Phi$,
we denote by
\ben
\delta F(\Phi) = \frac{\d}{\d t} F(\Phi_t) \mid_{t=0} = \left( \int_M
\delta \Phi(x) \frac{\delta}{\delta \Phi(x)} \right) F(\Phi)
\een
the variational derivative, along the ``vector field''
$\delta \Phi = \frac{d}{dt}\Phi_t \mid_{t=0}$ on field
space\footnote{In the integral, we
mean the ``left'' variational derivative when anticommuting quantities
are involved.}.
%Vector fields may be decomposed into a bosonic and fermionic
%part according to the ${\mathbb Z}_2$ grading of the coefficient algebra
%$A=A_+ \oplus A_-$.
We also define the commutator
\ben
[\delta_1, \delta_2] \Phi = \frac{\d^2}{\d t\d s}
[(\Phi_t)_s-
%(-1)^{\delta_1 \delta_2}
(\Phi_s)_t]
\mid_{t=s=0} \, .
\een
%of two vector fields on field space, where $(-1)^\delta$ is $-1$ for
%odd, and $+1$ for even elements.
For example, if $f_t$ is a 1-parameter
family of diffeomorphisms on $M$ generated by a vector field $X$,
and $\Phi_t = f^*_t \Phi$, then
\ben
\delta \Phi = {\pounds_X} \Phi \,
\een
and the commutator $[\delta_1,\delta_2]$ of variations associated
with two vector fields $X_1, X_2$ is given by
the vector field (on field space) ${\pounds_{[X_1,X_2]}} \Phi$, i.e.,
the variation associated with  the commutator of the spacetime vector fields.

The variation of $\LL$ can always be written in the
form
\ben
\delta \LL(\Phi) = \E(\Phi) \delta \Phi + \d \bt(\Phi, \delta \Phi) \, ,
\een
where $\E$ are the equations of motion (``Euler-Lagrange equations''),
and where $\bt$ is the $(d-1)$-form corresponding to the boundary term
that would be obtained if the variation of $\LL$ were
done under an integral. The (dual) symplectic current of the theory
$\bomega$ is defined
as the second anti-symmetrized variation
\ben\label{wdef}
\bomega(\Phi, \delta_1 \Phi, \delta_2 \Phi) =
\delta_1 \bt(\Phi, \delta_2 \Phi) -
%(-1)^{\delta_1\delta_2}
\delta_2 \bt (\Phi, \delta_1 \Phi) -\bt(\Phi, [\delta_1, \delta_2]
\Phi) \, . \een It is a simple consequence of the above definitions
that, if the fields $\Phi$ are solutions to the equations of motion,
and the variations $\delta_1 \Phi$ and $\delta_2 \Phi$ are solutions
to the linearized equations of motion, then $\bomega$ is closed, $\d
\bomega = 0$. Given a $d-1$-dimensional submanifold $\Sigma$ of $M$,
one defines the associated symplectic form by \ben
\sigma_\Sigma(\Phi; \delta_1 \Phi, \delta_2 \Phi) \equiv \lim_{K
\uparrow M} \int_K \bomega( \Phi; \delta_1 \Phi, \delta_2 \Phi) \, ,
\een where $K$ is an increasing sequence of compact subsets of
$\Sigma$ whose union is $\Sigma$. If $\Sigma$ is not compact,
suitable boundary conditions have to imposed on the fields in order
to ensure that the symplectic form is finite. One might also have to
impose restrictions on the possible way of choosing the sequence of
compact sets $K$ to ensure that the limit is unique. If
$\sigma_\Sigma$ converges, then it is independent under compact
cobordant variations of the surface $\Sigma$ in the sense that, if
$K$ is a compact subset of $M$ such that $\partial K = \Sigma_1
\sqcup (-\Sigma_2)$, then $\sigma_1 = \sigma_2$ for ``on-shell
variations'', i.e., $ \delta_1 \Phi$ and $\delta_2 \Phi$ satisfying
the linearized equations of motion and boundary conditions. Note,
however, that this need not be the case for non-compact cobordisms.
In that case, the independence of $\sigma_\Sigma$ from the slice
delicately depends on the asymptotic conditions imposed on the
fields. For example, it fails in general relativity when $\Sigma_1,
\Sigma_2$ are asymptotically null surfaces asymptoting to different
``cross sections'' at infinity, which is the situation considered in
the context of asymptotically flat spaces at null infinity.

A diffeomorphism preserving the boundary conditions is called an
``asymptotic symmetry''. Since diffeomorphisms leave the Lagrangian
invariant, one may ask whether one can define ``conserved charges''
in the phase space of the theory corresponding to the asymptotic
symmetries. They should be defined as the Hamiltonian generators
conjugate to a symmetry variation vector field $\delta \Phi =
\pounds_X \Phi$, i.e., they should satisfy \ben\label{dhx} \delta
{\mathcal H}_X = \sigma_\Sigma(\delta \Phi, {\pounds_X \Phi} ) =
\int_\Sigma \bomega(\Phi, \delta \Phi, {\pounds_X} \Phi) \quad
\forall \delta \Phi \quad {\rm on \,\, shell} \, . \een The
existence of the $\H_X$ is not automatically guaranteed by the
diffeomorphism invariance of the Lagrangian, but depends on the
precise choice of the boundary conditions and of $\Sigma$ in a
subtle way.
 To analyze the question of existence, consider (following~\cite{Iyer1994,Lee1989,Zoupas1998}) the consistency relation resulting from the fact that the second
anti-symmetrized second variation of any quantity should vanish, so
we should have
\ben\label{C}
\delta_1(\delta_2 \H_X) -
%(-1)^{\delta_1\delta_2}
\delta_2(\delta_1 \H_X) - [\delta_1,\delta_2] \H_X = 0
\quad
\text{for all $\delta_1, \delta_2$.}
\een
Here, the last term takes into account the possibility
of having non-commuting variations; it vanishes
when $\delta_1, \delta_2$ are commuting variations
(e.g., corresponding to shifts of the  phase space coordinates).
As a consequence of eq.~\eqref{wdef}, we have the relation
%(for even vector fields of field space)
\bena
0&=&
\delta_{1} \bomega(\Phi, \delta_2 \Phi, \delta_{3} \Phi)
+\delta_{2} \bomega(\Phi, \delta_3 \Phi, \delta_{1} \Phi)
+\delta_3 \bomega(\Phi, \delta_1 \Phi, \delta_{2} \Phi)
\nonumber\\
&+& \bomega(\Phi, \delta_1\Phi, [\delta_2, \delta_3] \Phi)
+ \bomega(\Phi, \delta_2 \Phi, [\delta_3, \delta_1] \Phi)
+ \bomega(\Phi, \delta_3 \Phi, [\delta_1, \delta_2] \Phi) \, .
\eena
%and a similar relation for vector fields of arbitrary bose/fermi character.
Using this with $\delta_3 \Phi = \pounds_X \Phi$,
one finds that the consistency condition~\eqref{C}
can be expressed as
\bena
0 &=& \int_\Sigma {\pounds_X} \bomega(\Phi, \delta_1 \Phi, \delta_2 \Phi) \\
  &=& \int_{\partial \Sigma}
X \cdot \bomega(\Phi, \delta_1 \Phi, \delta_2 \Phi) \, .
\eena
In the second line, we have used Stokes theorem, together with the
the standard differential forms
identity ${\pounds_X} \bomega = \d (X \cdot \bomega) +
X \cdot \d \bomega$, and the fact that $\bomega$ is closed on shell.
The notation ``$\partial \Sigma$'' indicates any interior actual
boundaries, as well as possibly a boundary at ``infinity''. In the
examples below, we will attach such a boundary $\partial M = \I$ to our
spacetime manifold, and $\Sigma$ will be a $(d-1)$-dimensional submanifold
which meets $\I$ in a cross section, $C$.
A sufficient condition for the consistency
condition to hold is that
\ben
X \cdot \bomega \restriction_{\partial
\Sigma} = 0 \, .
\een
 The vector fields $X$ for which the
above manipulations are expected to give conserved charges are those
vector fields whose asymptotic values at conformal infinity leave
the boundary conditions imposed upon the fields $\Phi$ invariant.
Typically, the restrictions of those vector fields to the boundary
span the tangent space of $\I$. Thus, a sufficient condition for the
consistency condition~\eqref{C} to hold is that \ben \label{cons}
\bomega \restriction_{\I} = 0 \, . \een Whether this condition holds
depends on both (a) the Lagrangian and (b) the precise form of the
boundary conditions. We will
 analyze this issue in some example theories below.
When eq.~(\ref{C}) does hold,
the conserved charges can be obtained from eq.~\eqref{dhx} as follows.
Choose any path $\Phi_t$ of solutions
from a canonically fixed reference solution
$\Phi_0$ to $\Phi=\Phi_1$. Define
\ben
\H_X(\Phi) = \H_X(\Phi_0) + \int_0^1 \delta \H_X(\Phi_t) \, dt \, .
\een
where $\delta \Phi$ is the vector field along the path, and
where $\H_X(\Phi_0)$ is defined arbitrarily. Then it follows
from the consistency condition~\eqref{C} that the
definition of $\H_X$ is unchanged under smooth deformation of the
path, i.e., it only depends (possibly) on the homotopy class of the
path in the space of solutions of the theory. Whence, if this space
is for example simply connected, then the definition of $\H_X$ is unique.
On the other hand, if it is not simply connected, the $\H_X$ might be
multi-valued and might consequently
be defined only on the covering space of the
covariant phase space.

 Furthermore, if eq.~\eqref{cons} holds (implying~\eqref{C}),
then $\H_X$ will automatically
be independent of the choice of the surface $\Sigma$, i.e., the cut
$C=\partial \Sigma$ where $\Sigma$ hits the boundary.
This
follows straightforwardly from Stokes theorem, and
the fact that $\d \bomega = 0$ on shell.

\medskip
\noindent

As we now describe in some detail, these considerations admit a more
or less straightforward generalization to symmetries parametrized by
arbitrary tensor fields. Let us assume that there exists a symmetry
$\b_\eta$ of the Lagrangian $\LL$ for each choice of some spinor or
tensor field $\eta$. In other words, assume that for each $\eta$
there exists a vector field $\b_\eta\Phi$ on field space such that
 \ben \b_\eta\LL(\Phi) = \d \B_\eta(\Phi), \een
where $\B_\eta$ is a $(d-1)$-form that is locally constructed out of
the fields and $\eta$. This includes in particular the case of
infinitesimal diffeomorphism invariance, where $\eta$ is simply
given by a vector field $\eta =X^\mu \partial_\mu$, and where $\b_X
\Phi= {\pounds_X} \Phi$, and $\B_X = X \cdot \LL$. More generally,
it covers e.g. local gauge transformations, and common fermionic
symmetries such as supersymmetry. The precise form of the symmetry
is unimportant for our present discussion. For definiteness, let us
assume that $\eta$ is an (anti-commuting) spinor field; this is the
case
 for  supersymmetry,
which is the context of primary interest in this paper.

We would like to define
corresponding Hamiltonian generators in the phase space of the
theory by
\ben\label{dh}
\delta {\mathcal H}_\eta =
\int_\Sigma \bomega(\Phi, \delta \Phi, \b_\eta \Phi)
\, .
\een
(Recall that, by the general definition, the symmetry variation of
a functional $F$ on field space
is given by $\b_\eta F(\Phi) =
\frac{d}{dt} F(\Phi + t \b_\eta \Phi) |_{t=0}$).
To see whether the generator exists, we must analyze the consistency
of this equation in the same fashion as we did above for diffeomorphisms, by
taking a second antisymmetrized variation, see eq.~\eqref{C}. We now get the
consistency condition
\ben
\label{cons2}
0=\int_\Sigma \b_\eta \bigg[ \bomega(\Phi, \delta_1 \Phi,
\delta_2 \Phi) \bigg] \, ,
\een
for all solutions $\Phi$, and all on-shell variations. Here
$\b_\eta$ is defined to act on variations $\delta \Phi$
as the linearized symmetry transformation, i.e. the transformation
on the tangent space of field space induced by $\b_\eta$. Since
$\d \bomega = 0$ it follows that
\ben
0=\d \left\{
\b_\eta \bigg[ \bomega(\Phi, \delta_1 \Phi,
\delta_2 \Phi) \bigg]
\right\} \, ,
\een
for all $\eta$. By the fundamental lemma of the calculus of
variations\footnote{Fundamental Lemma (see e.g., \cite{Wald1988}):
Let $\alpha[\psi]$ be a $p$-form on
an $n$-dimensional manifold $M$, with $p<n$, depending on a set of fields $\psi$ on $M$, such
that $\d \alpha[\psi] = 0$ for {\em all} $\psi$,  and such that $\alpha=0$ when the non-dynamical fields in $\psi$ vanish. Then $\alpha[\psi] = \d \beta[\psi]$
for some globally defined form $\beta$.
Note that it is crucial that the relation
$\d \alpha[\psi] = 0$ holds for all $\psi$. Otherwise, one can only infer
the existence of $\beta$ locally, but possibly not globally if the
topology of $M$ is non-trivial.}~\cite{Wald1988}
it consequently follows that there exists a
$(d-2)$-form $\Y_\eta$, such that
\ben
\label{3}
\b_\eta \bigg[ \bomega(\Phi, \delta_1 \Phi,
\delta_2 \Phi) \bigg] =
\d \Y_\eta(\Phi, \delta_1 \Phi, \delta_2 \Phi) \, .
\een
Therefore, by Stokes theorem, a sufficient condition for the consistency
of eq.~\eqref{cons2} is that
\ben
0=\int_{\partial \Sigma} \Y_\eta \, .
\een
This condition is automatically satisfied if
\ben
\Y_\eta \restriction_\I = 0 \, .
\een
Thus, if this equation holds (and in particular, if the boundary
conditions are such that $\Y_\eta$ is actually finite at $\I$),
then charges $\H_\eta$ will exist, and will be conserved, i.e.,
do not depend on the particular surface $\Sigma$ chosen in the
definition.

A sufficient condition for $\int_{\partial \Sigma}
\Y_\eta \restriction_\I = 0$
can be derived as follows. First, assume that the boundary
conditions are such that the consistency condition
$\bomega \restriction_\I = 0$ holds, which, as we showed above,
is a sufficient condition for the bosonic asymptotic symmetries to
exist.
Now, suppose that  the $\eta$-variations leave
the symplectic form invariant on $\I$.  Then by (\ref{3})
$\d \Y_\eta \restriction \I
 = 0$. Consequently, the integral $\int_C \Y_\eta(\Phi; \delta_1 \Phi, \delta_2 \Phi)$ does not depend
upon the cut $C$ at $\I$. Consider now perturbations $\delta \Phi$ in this
expression which are compactly supported on the particular slice $\Sigma$, i.e., that vanish
on the particular cut $C = \partial \Sigma$. Then the integral
vanishes for that cut, and hence for any cut. Thus, the consistency
condition is fulfilled for any variation for which the ``initial data'' on $\Sigma$ has compact support.
In contexts with timelike conformal boundary (such as asymptotically AdS spacetimes), this implies that
this integral vanishes for arbitrary perturbations (satisfying the linearized
form of the boundary conditions). Thus, in this situation it follows that
{\em the  charges $\H_\eta$ will exist and will be conserved if}
\begin{enumerate}
\item The pull back of the symplectic form $\bomega \restriction_\I$ to
$\I$ vanishes identically under the assumed boundary conditions.
\item The symmetry leaves the boundary conditions invariant in the
  sense that if is a field configuration $\Phi$ satisfying
  the boundary conditions, then $\b_\eta \Phi$ satisfies
the linearized boundary conditions.
\end{enumerate}

\medskip
\noindent

It is a general feature of symmetries $\b_\eta$ parametrized by an
arbitrary tensor or spinor field $\eta$ (such as infinitesimal
diffeomorphisms, gauge symmetries, or supersymmetry transformations) that the
generating conserved charges can be expressed in terms of
surface integrals at infinity.
In the present formalism, one may derive this fact by generalizing the
argument given by Wald et
al.~\cite{Iyer1994,Iyer1995,Lee1989,Zoupas1998} for diffeomorphisms.

As with diffeomorphisms, the starting point is the Noether current, defined by
\ben
\J_\eta(\Phi) = \bt(\Phi, \b_\eta \Phi) - \B_\eta \, .
\een
On shell (i.e., when the equations of motion hold), this current is
conserved,
\ben
\d \J_\eta = \d \bt(\phi, \b_\eta \Phi) - \d \B_\eta = \b_\eta \LL - \d \B_\eta - \E \cdot \b_\eta \Phi = 0 \, .
\een
By the ``fundamental lemma'', since this is true for {\em all}
$\eta$, it follows that there exists a $(d-2)$-form,
$\Q_\eta$, which is locally constructed out of the fields,
such that
\ben\label{noether}
\d \Q_\eta = \J_\eta \, .
\een
We refer to $\Q_\eta$ as the ``fermionic'' Noether charge.
The integrand of eq.~\eqref{dh} may then be written as
\ben
\bomega(\Phi, \delta \Phi, \b_\eta \Phi)
=\delta \d \Q_\eta(\Phi) + \delta \B_\eta(\Phi)
- \b_\eta \{ \bt(\Phi, \delta \Phi) \} \, ,
\een
where the symmetry variation in the last expressions
acts on $\delta \Phi$ by the linearized symmetry transformations.
Since $\b_\eta \Phi$ is a solution to the linearized
equations of motion, we have $\d \bomega(\Phi,\delta \Phi, \b_\eta \Phi) =
0$, so by the fundamental Lemma of variations $\bomega(\Phi,\delta \Phi,
\b_\eta \Phi)$ is exact, and there is a
$(d-2)$-form $\A_\eta(\Phi, \delta \Phi)$ such that
\ben
\label{Adef}
\d \A_\eta(\Phi, \delta \Phi) =
\b_\eta \{ \bt(\Phi, \delta \Phi) \} - \delta \B_\eta(\Phi)  \, ,
\een
for any solution $\Phi$ to the equations of motion, and any
solution $\delta \Phi$ to the linearized equations.
Therefore, we have
\ben\label{Hdef}
\delta {\mathcal H}_\eta =
\int_{\partial \Sigma} (\delta \Q_\eta - \A_\eta) \, ,
\een
thus expressing $\delta \H_\eta$
as a boundary integral as promised.
For ``off shell'' variations $\delta \Phi$,
$\delta \H_\eta$ is not in general
a boundary integral, but one can show (see Appendix) that it can
always be written as the above boundary integral plus an integral
over $\Sigma$ of suitably defined constraints associated with the 
symmetry $s_\eta$.

The consistency requirement $\delta_1(\delta_2 \H_\eta)
%- (-1)^{\delta_1 \delta_2}
-\delta_2(\delta_1 \H_\eta) - [\delta_1,\delta_2] \H_\eta=0$
is now equivalent to
\ben\label{CC}
0 = \delta_1 \A_\eta(\Phi, \delta_2 \Phi)
-
%(-1)^{\delta_1 \delta_2}
\delta_2 \A_\eta(\Phi, \delta_1 \Phi)
+ \A_\eta(\Phi, [\delta_1, \delta_2]\Phi) \quad {\rm on}\,\,
\partial \Sigma \,
\een
for all variations $\delta_1 \Phi, \delta_2 \Phi$ satisfying
the linearized equations of motion.
Whether this condition holds again depends on the theory under
consideration, and on the choice of the boundary conditions.
Assuming that the consistency condition~\eqref{CC}
holds,
 the equation for $\delta \H_\eta$ can be integrated as above
in the case of ``bosonic'' symmetries $X$, and,
eq.~\eqref{Hdef}, $\H_\eta$ is expressed by a boundary integral on
shell. An off-shell expression for $\H_\eta$ is given
in the appendix [see eq.~\eqref{offshell}], 
where also various issues related to the 
uniqueness and gauge invariance of $\H_\eta$ are discussed in detail. 
%The off-shell formula also involves a bulk integral over $\Sigma$
%involving the constraints of the theory associated with the symmetry 
%$s_\eta$. 

Let us now consider the special case of bosonic symmetries associated
with infinitesimal diffeomorphisms, $\b_X \Phi = \pounds_X \Phi$,
with $X$ a suitable vector field so that we may show explicitly that
the above procedure reduces to the ``covariant phase space method'' of
Wald et al. In that case, because
\ben
\b_X \LL = \pounds_X \LL = \d (X \cdot \LL) + X \cdot \d \LL = \d (X \cdot
\LL) \, ,
\een
we have $\B_X = X \cdot \LL$. On shell, we therefore have
$\delta \B_X = X \cdot \delta \LL =
X \cdot \d \bt$, and thus, from the defining relation for
$\A_X$, we get
\ben
\A_X = X \cdot \bt \, .
\een
On shell, we therefore have
\ben
\delta \H_X = \int_C \delta \Q_X - X \cdot \bt
\een
for symmetries associated with infinitesimal diffeomorphisms. This is
the formula derived by Wald et al.~\cite{Iyer1994,Iyer1995,Zoupas1998}.

\section{${\mathcal N}=1$ minimal Sugra with cosmological constant}
\label{minimal}

We now illustrate the above algorithm in the example theory of
$\N=1$ supergravity with a negative cosmological
constant~\cite{Townsend1977} (treated also in the paper of Henneaux
and Teitelboim~\cite{Henneaux1985} using a different technique). The
dynamical fields in this theory are a metric and an anti-commuting
spin-(3/2) Majorana field, \ben \Phi = (g_{\mu\nu}, \psi_\mu) \, .
\een The Lagrange 4-form is given by \ben\label{min} \LL =
\left[\frac{1}{2} R + \epsilon^{\lambda \rho \mu \nu} \bar
\psi_\lambda \Gamma \gamma_\mu D_\nu \psi_\rho + 3\ell^2 +
\frac{i}{2}\ell \bar \psi_\lambda \gamma^{[\lambda} \gamma^{\rho]}
\psi_\rho \right] \beps \, , \een where
$\beps=\epsilon_{\mu\nu\sigma\rho}$ is the volume 4-form determined
by the metric, and where \ben \Gamma = (i/4!)
\epsilon^{\mu\nu\sigma\rho} \gamma_\mu \gamma_\nu \gamma_\sigma
\gamma_\rho \een is the analogue of $\gamma_5$ in curved spacetime.
The derivative operator $D_\mu$ is defined by
 \ben\label{dm} D_\mu =
 \nabla_\mu + \frac{i}{8}[\gamma^\sigma,\gamma^\rho](\bar \psi_\sigma
 \gamma_\mu \psi_\rho + \bar \psi_\mu \gamma_\sigma \psi_\rho - \bar
 \psi_\mu \gamma_\rho \psi_\sigma) \, ,
 \een
where $\nabla_\mu =
\partial_\mu + C_\mu(g)$ is the spin connection of the metric. The
quantity $R$ is the scalar curvature formed from the curvature
2-form of  $D_\mu$,
 \ben
 R_{\mu\nu} = [D_\mu, D_\nu]
%\d \omega^{ij} + \omega^{ik} \wedge \omega_{k}{}^j \, .
 \een
and $\ell$ is a constant.

When the spinor field
is zero, the Lagrangian has the standard form of the Einstein
action,
with a negative cosmological constant $\Lambda = -3\ell^2$. Exact
AdS spacetime with topology $\mr^3 \times \mr$ and line element
\ben
\d s^2_0 = -(1+\ell^2/r^2)\d t^2 + (1+\ell^2/r^2)^{-1} \d r^2 + r^2
(\d \theta^2 + \sin^2 \theta \, \d \varphi^2)
\een
is an exact solution of this theory, and it appears reasonable to demand as
a boundary condition that
the metric of a general solution should asymptotically
approach that of exact AdS for large distances. A convenient and
elegant framework for formulating this requirement is that of
conformal infinity. This is motivated by the fact that the AdS metric
can be written as
\ben\label{global}
\d s^2_0 = \frac{\ell^2}{\Omega^2} \bigg[\d\Omega^2 - (1 + \Omega^2/2)\d
t^2 +
(1-\Omega^2/2)
(\d \theta^2 + \sin^2 \theta \, \d \varphi^2)
 \bigg] = \Omega^{-2}
\d\tilde s^2_0
\een
where $\Omega$ is related to $r$ by a simple coordinate transformation
satisfying $\Omega \sim r^{-1}$ for large $r$. Thus, the AdS metric is
conformally related to the ``unphysical metric'' $\d\tilde s^2_0$. The
point is now that the
unphysical metric can be smoothly extended to a manifold with boundary
$\tilde M = M \cup \I$, where the infinity $\partial M = \I \cong \mr
\times S^2$
is represented by the points labelled by $\Omega = 0$. The metric
induced on $\I$ is that of the Einstein static universe (ESU),
$\ell^2[-\d t^2 + \d \theta^2 + \sin^2 \theta \, \d \varphi^2]$.
This motivates one to impose the following asymptotically AdS
boundary conditions on the metric:

\paragraph{Boundary conditions on $g_{\mu\nu}$:}
\begin{enumerate}
\item One can attach a boundary $\I = S^2 \times \mr$ to the spacetime
manifold $M$ in such a way that $\tilde M = M \cup \I$ is a manifold
with boundary. $\I$ is called the ``conformal infinity'', or scri.

\item On $\tilde M$, there is a smooth field $\Omega$, with the properties
that $\Omega$ is a defining function for $\I$ (i.e., vanishes there and
has non-vanishing gradient), and such that
\ben
\tilde g_{\mu\nu}=\Omega^2 g_{\mu\nu}
\een
is smooth at scri. If
\ben
\tilde n^\mu = \ell^{-1} \tilde g^{\mu\nu} \tilde \nabla_\nu \Omega \, .
\een
then we impose that $\tilde n^\mu$ is a unit normal to the boundary
$\I$ of the unphysical spacetime $(\tilde M, \tilde g_{\mu\nu})$, and
we impose that the induced metric on $\I$ is isometric to that of the
Einstein static universe $\mr \times S^2$.
We further impose that the extrinsic curvature of $\I$,
$\tilde K_{\mu\nu} = \tilde \nabla_{(\mu} \tilde n_{\nu)} = 0$ at
$\I$, and that $\Omega^{-1} \tilde K_{\mu\nu}$ is smooth. These
conditions are equivalent to demanding that
\ben
\d s^2 = \frac{\ell^2}{\Omega^2} \Bigg[ \d \Omega^2 -
%\left(1 - \frac{1}{2} \Omega^2 \right) \,
\d t^2
+
%\left(1 - \frac{1}{2} \Omega^2 \right) \,
\d \theta^2 + \sin^2 \theta \, \d \varphi^2
\Bigg] + O(\Omega^0)
%\nonumber \\
%&& + \Omega^3 \tilde b_{\mu\nu} \, \d x^\mu \d x^\nu
%+ O(\Omega^4) \Bigg] \, .
\een
Here, the notation $t_{\mu\nu \dots \sigma} = O(\Omega^a)$ for a
tensor field means that $\Omega^{-a} t_{\mu\nu \dots \sigma}$
can be extended to a smooth tensor field $\tilde t_{\mu\nu \dots
  \sigma}$ on $\tilde M$.
\end{enumerate}
As we will see below, Einstein's equations imply more stringent
constraints upon the asymptotic form of the metric in addition to the
above boundary conditions.
From now on, we will set
\ben
\ell = 1 \, .
\een

An asymptotic (bosonic) symmetry is a diffeomorphism $f$ having the
property that $f^* \d s^2$ satisfies our asymptotic conditions whenever the
metric $\d s^2$ does. Evidently, if $f$ acts like a conformal isometry on
the ESU with conformal factor $k^2$, then we may change $\Omega$ to
$k\Omega$, and the unphysical metric $\d \tilde s^2$ associated
with that new conformal factor will satisfy our asymptotic
conditions. Thus, the group of asymptotic symmetries contains the
conformal isometry group $O(3,2)$ of the Einstein static universe.
It is customary to consider as the true physical symmetry group the
group of all diffeos $f$ as described above, factored by the the group
of ``pure gauge'' transformations, which is the group of all $f$
leaving $\I$ pointwise fixed. With this definition, the group of
asymptotic symmetries is isomorphic to $O(3,2)$.

Since we are in $d=4$ dimensions, one may also impose boundary
conditions \cite{IW} which have slower fall-off near $\I$. However,
AdS/CFT arguments \cite{LP1,LP2,MR} suggest that such boundary
conditions are related to ``dual'' gravitational theories in
spacetimes with compact Cauchy surfaces.  As a result, we expect
that with such slower fall-off boundary conditions all asymptotic
symmetries will be gauge.  This result is also suggested by the
analysis of \cite{Dmitry}.  In particular, we expect that such
boundary conditions will not lead to interesting conserved
charges\footnote{For $d=4,5,6$, it is
  possible \cite{IW} to introduce a further class of boundary
  conditions (termed `hybrid boundary conditions' in \cite{MR} by
  introducing a preferred direction field on $\I$.  Since this
  preferred direction field will break supersymmetry, we will not
  explore such boundary conditions here.  }, and we will not consider them.

The Lagrangian $\LL$ is diffeomorphism invariant. It is also
invariant under supersymmetry transformations of the form
\bena
\b_\eta g_{\mu\nu} &=& - 2i \bar \eta \gamma_{(\mu} \psi_{\nu)} \\
%\b_\eta \omega_{\mu \, ij} &=& S_{\mu \, ij} -
%S_{\tau [i}{}^\tau e_{j] \mu}\\
\b_\eta \psi_\mu &=& \left(D_\mu + \frac{1}{2} \gamma_\mu \right)
\eta \, ,
\eena
%where
%\ben
%S_i{}^{\mu\nu} = \epsilon^{\lambda\rho\mu\nu}\bar \eta \Gamma \gamma_i
%\widehat \nabla_\lambda \psi_\rho \, \qquad
%\widehat D_\mu = D_\mu + \frac{i}{2\ell} \gamma_\mu \, ,
%\een
in the sense that $\b_\eta \LL = \d \B_\eta$ for some 3-form $\B_\eta$,
for any Majorana field $\eta$. We would like to construct corresponding
generators $\H_\eta$ on the covariant phase space. According to the
general recipe presented in the previous section, the existence of
such generators will follow if we impose boundary conditions on the
fields $(g_{\mu\nu}, \psi_\mu)$
such that  (1) and (2) hold.
We have already chosen boundary
conditions for the metric $g_{\mu\nu}$,
but we have as yet not chosen any boundary conditions on the spinor
fields $\psi_\mu$. We will now choose them in such a way that
(1) and (2) are satisfied.

Condition
(1)  requires us to know the symplectic current of the
theory. It is given by
\ben
\bomega_{\mu\nu\sigma} = \epsilon_{\mu\nu\sigma\rho} \, *\!\bomega^\rho \, ,
\een
where
\bena
*\bomega^\mu &=& P^{\alpha \beta \gamma \tau \sigma \mu}
[\delta_1 g_{\alpha\beta} \delta_2 C_{\gamma \, \tau \sigma} -
(1 \leftrightarrow 2)]
\nonumber\\
&+& \epsilon^{\mu\alpha\beta\gamma} [\delta_1 \bar \psi_\alpha
\Gamma \gamma_\beta \delta_2 \psi_\gamma + \frac{1}{2}
\delta_1 g_\beta{}^\tau
\bar \psi_\alpha \Gamma \gamma_\tau \delta_2 \psi_\gamma
-(1 \leftrightarrow 2)] \, ,
\eena
and where
\ben
P^{\alpha\beta\gamma\tau\sigma\mu} = -\frac{1}{2}
g^{\alpha\beta}g^{\gamma\tau}g^{\mu\sigma}
+ \frac{1}{2}
g^{\alpha\mu}g^{\gamma\tau}g^{\beta\sigma}
+ \frac{1}{2}
g^{\alpha\gamma}g^{\beta\tau}g^{\mu\sigma} \, .
\een
In these formulae, $\delta C_{\mu \nu \sigma}$ is a shorthand for
\ben
\delta C_{\mu \, \nu\sigma} = \delta C_{\mu \, \nu\sigma}(g) -
\frac{i}{2}\delta (\bar
\psi_\nu \gamma_\mu \psi_\sigma + \bar \psi_\mu \gamma_\nu \psi_\sigma - \bar
\psi_\mu \gamma_\sigma \psi_\nu) \,
\een
where $\nabla_\mu = \partial_\mu + C_\mu(g)$ is the spin connection of
$g_{\mu\nu}$.
% and where $\pm (1 \leftrightarrow 2)$ means that we
%chose $\pm$ according to the bose/fermi parity of the variations.

It can be seen that the
boundary conditions imposed upon $g_{\mu\nu}$
imply that the bosonic contribution to the symplectic form is
finite at $\I$, and that the corresponding contribution to the
symplectic flux vanishes.
If the spinor field $\psi_\mu$ is of order $\Omega^{1/2}$,
then the contribution to the symplectic form from the
spinor contribution to $\delta C_{\mu \, \nu\alpha}$ is of
order $\Omega^2$ and gives no contribution to the symplectic flux.
The finiteness of the contribution quadratic in the
variation of the spin 3/2-field to the symplectic form
at infinity requires the spinor field $\delta \psi_\mu$
to fall off at least as fast as $\Omega^{1/2}$. From
a geometric point of view, it is  thus natural to require that the field
$\Omega^{-1/2} \psi_\mu$ be finite and smooth
at $\I$. The vanishing of the spinor contribution to the symplectic
flux through $\I$, given by
 \ben
 F_{spin-3/2} = \int_{\I}
\overline{\delta_1 \psi_{\mu}}
\gamma_\nu \Gamma \delta_2 \psi_{\sigma} \,
 \d x^\mu \wedge \d x^\nu \wedge \d x^\sigma ,
 \een
is not ensured however by merely requiring that $\Omega^{-1/2}
\psi_\mu$ be smooth at $\I$. A sufficient condition is obviously that
the pull-back to $\I$ of the integrand vanishes,
\ben\label{vani}
\overline{\delta_1 \psi_{[\mu}}
\gamma_\nu \Gamma \delta_2 \psi_{\sigma]} \, \restriction \I = 0  \, .
\een
We now fix a boundary condition on $\psi_\mu$ which
does precisely this.

\paragraph{Boundary condition for $\psi_\mu$:}
\ben
\psi_\mu = O(\Omega^{3/2}) \, .
\een
%for some asymptotic Killing spinor $\eta$.
%\Omega^{3/2} \left[ \Omega^{-1} \tilde n_\mu \tilde P_{-}
%\tilde U + \tilde P_+ \tilde V_\mu + O(\Omega) \right] \, ,
%\een
%where $\tilde V_\mu$ is finite at $\I$, and where $\tilde U$ is
%a spinor field as above in eq.~\eqref{gauge}.

We next need to see whether our boundary conditions on $(g_{\mu\nu},
\psi_\mu)$ are compatible with supersymmetry.
On exact AdS-spacetime, one can choose a basis of covariantly constant
Majorana spinors, i.e., global solutions to the equation
\ben
\label{bgspinor}
\left(
{\stackrel{\circ}{\nabla}}_\mu + \frac{1}{2}
{\stackrel{\circ}{\gamma}}_\mu
\right) \eta = 0 \, ,
\een
where the circle above derivative operator and gamma-matrix
indicate that the quantities associated with exact AdS spacetime
are meant in this equation. Defining
the ``chiral projections''
 \ben
 \tilde P_{\pm} =
 \frac{1}{2}(1 \pm \tilde n^\mu \tilde \gamma_\mu),
 \een
(so that $\tilde P^2_\pm = \tilde P^{}_\pm, \tilde P_+^{} \tilde
P^{}_-=0$ everywhere), one may show that, in any spacetime, one has
\ben\label{gauge0} \left( \nabla_\mu + \frac{1}{2} \gamma_\mu
\right)(\Omega^{-1/2} \tilde \psi) = \Omega^{1/2}\left( \tilde
\nabla_\mu + \Omega^{-1} \tilde \gamma_\mu \tilde P_- \right) \tilde
\psi \, , \een for any spinor field $\psi$. Thus, taking $\psi =
\eta$, and using the Killing-spinor equation, we must have that
$\tilde P_- \tilde \eta \restriction \I = 0$. In fact, one can show
that \ben\label{gauge} \tilde \eta = \Omega^{1/2} \eta = \tilde P_+
\tilde U + \frac{1}{2} \Omega \tilde P_- \tilde U + O(\Omega^2) \een
for certain spinor fields $\tilde U$~\cite{Henneaux1985,BF} which
are smooth at scri. On a generic spacetime that is only
asymptotically AdS, there do not exist any spinors satisfying
eq.~(\ref{bgspinor}) However, using the asymptotic form of the
metric and the fact that $\eta$ is a Killing spinor of AdS, one may
show
 \bena\label{var} \Omega^{-3/2}  \left( \nabla_\mu +
 \frac{1}{2\ell} \gamma_\mu \right) \eta &=&  \Omega^{-1} \tilde
 \gamma^\nu (\Omega^{-1} \tilde
 K_{\mu\nu}) \tilde P_+ \tilde U \non\\
 && - \tilde \gamma^\nu (\Omega^{-1} \tilde
 K_{\mu\nu}) \tilde P_- \tilde U
 + \tilde \gamma^\nu \tilde \nabla_\nu (\Omega^{-1} \tilde
 K_{\mu\alpha}) \tilde \gamma^\alpha \tilde P_- \tilde U \non\\
 && - \frac{1}{2} \tilde \nabla^\nu (\Omega^{-1} \tilde K_{\mu\nu})
 \tilde P_+ \tilde U + \dots
%
%-\frac{1}{2} a
%\Omega^{-1} \tilde n_\mu \tilde P_- \tilde U\\
%&& +\frac{1}{4\ell} a \tilde n_\mu \tilde P_+ \tilde U\\
%&& -\frac{3}{2} b_{\mu\nu} \tilde P_+ \tilde \gamma^\nu \tilde U
%+ \frac{3}{16} b_{\alpha\beta} \tilde h^{\alpha\beta}
%\tilde P_+ \tilde \gamma_\mu \tilde U \, ,
 \eena
where dots stands for terms of order $O(\Omega)$. In these formulae,
$\tilde K_{\mu\nu} = \tilde \nabla_{(\mu} \tilde n_{\nu)}$
is the extrinsic curvature of $\I$ with respect to
the unphysical metric. Note that our boundary conditions on the
metric require that $\Omega^{-1} \tilde K_{\mu\nu}$ is smooth at $\I$.
In deriving the above equation, the formula
\ben
\delta( \nabla_\mu \psi) = -\frac{i}{8} [\gamma^\alpha, \gamma^\beta]
(\nabla_\alpha \delta g_{\mu \beta} - \nabla_\beta \delta
g_{\alpha\mu}) \psi
\een
for the linearized spin connection was found to be useful.
%plus higher order terms. Here $a$ is the coefficient appearing in the
%$O(\Omega^0)$ piece in $g_{\mu\nu}$, while $b_{\mu\nu}$ stands for the
%$O(\Omega^1)$ coefficient appearing in an expansion of $g_{\mu\nu}$
%around the boundary. We may assume that $\Omega$ is chosen so that
%$b_{\mu\nu}$ is tangent to $\I$.

We can now analyze the question  of whether our boundary conditions
are left invariant by the supersymmetry transformations. Consider
first the variation $\b_\eta \psi_\mu$, with $\eta$ a Killing spinor
of exact AdS spacetime. From our boundary condition on $\psi_\mu$, we
know that
%% we know that
\ben
D_\mu - \nabla_\mu =  \frac{i}{8} [\gamma^\nu, \gamma^\sigma]
(\bar
\psi_\nu \gamma_\mu \psi_\sigma + \bar \psi_\mu \gamma_\nu \psi_\sigma - \bar
\psi_\mu \gamma_\sigma \psi_\nu) \,
 = O(\Omega^4).
\een
Thus, $(D_\mu + \frac{1}{2} \gamma_\mu) \eta$ is given by
eq.~\eqref{var}, up to higher order terms. However, as it stands,
it follows from~\eqref{var} only that
\ben\label{var1}
\left( D_\mu + \frac{1}{2} \gamma_\mu \right) \eta
= \Omega^{1/2} \tilde \gamma^\nu (\Omega^{-1} \tilde
K_{\mu\nu}) \tilde P_- \tilde U + O(\Omega^{3/2})
\een
and thus only that $\b_\eta \psi_\mu = O(\Omega^{1/2})$, instead of
our desired boundary condition $\b_\eta \psi_\mu = O(\Omega^{3/2})$.
However, we will now use the equations of motion to show that in fact
$\Omega^{-1} \tilde K_{\mu\nu} = 0$ on $\I$, so the desired boundary
condition follows for $\b_\eta \psi_\mu$. For this, it is convenient
to make a decomposition
\ben
\tilde g_{\mu\nu} = \tilde n_\mu \tilde n_\nu + \tilde h_{\mu\nu}
\een
of the unphysical metric into ``Gaussian normal form'',
where $\tilde n^\mu = (\partial/\partial \Omega)^\mu$ is geodesic,
and where $\tilde h_{\mu\nu}(\Omega)$ is the induced metric on the
surfaces of constant $\Omega$. Let $T_{\mu\nu}$ be the matter stress
tensor, containing all terms in Einstein's equation involving the
spin-3/2 field, let $\tilde L_{\mu\nu} = T_{\mu\nu} - (1/3) g_{\mu\nu}
T$ and let ${\mathcal R}_{\mu\nu}$ be the Ricci tensor of the
metric $\tilde h_{\mu\nu}$ on the submanifolds of constant
$\Omega$. Then Einstein's equation may be
% decomposed as~\cite{marolf}
decomposed as\footnote{Here we follow the notation of~\cite{marolf}.
The recurrence scheme dates back at least to \cite{FG,Starob},
and has been much developed (see e.g. \cite{KS5,KS6} and
references therein) for the purposes of AdS/CFT.}
\bena\label{recursion}
\frac{\partial}{\partial \Omega} \tilde K_\alpha{}^\beta
&=& -\R_\alpha{}^\beta + \tilde K \tilde K_\alpha{}^\beta
+ 2\Omega^{-1} \tilde K_\alpha{}^\beta + \Omega^{-1} \tilde K
\delta_\alpha{}^\beta \non\\
&&
+ \tilde h_\alpha{}^\gamma \tilde h^{\beta\delta} \tilde
L_{\gamma\delta} + \frac{1}{2} \tilde L \delta_\alpha{}^\beta\\
\frac{\partial}{\partial \Omega} \tilde h_{\alpha\beta}
&=& -2\tilde h_{\beta\gamma} \tilde K_\alpha{}^\gamma \, .
\eena
Here $\partial/\partial \Omega$ should be understood geometrically
as $\pounds_{\tilde n}$.
The boundary condition on $\psi_\mu$ now implies that
$\tilde L_{\alpha\beta} = O(\Omega)$, from which it can
be seen to follow that even $\tilde K_{\mu\nu} = O(\Omega^2)$, and
not just $\tilde K_{\mu\nu} = O(\Omega)$, as required by our
boundary condition on the metric.
Thus, it follows from~\eqref{var1} that $\b_\eta \psi_\mu =
O(\Omega^{3/2})$, showing that the supersymmetry variation
of the spin-3/2 field satisfies the boundary conditions (on shell).

The supersymmetry transformation of the
metric under the above boundary conditions can
be checked immediately to be of order
\ben
\b_\eta g_{\mu\nu} = O(\Omega^0) \, .
\een
Thus, the supersymmetry variation of the metric
also satisfies our boundary conditions.

Of course, it remains to be seen that the boundary conditions are also
consistent with the field equations, in the sense that there is
a wide class of solutions. For the metric, we have
given an argument for this above, assuming that the stress tensor
is vanishing sufficiently rapidly.
Thus, a wide class of solutions with $\psi_\mu=0$
is allowed. For the spin-3/2 field, the
issue may e.g. be studied via a
linearized perturbation analysis.  We shall not undertake a detailed study of such perturbations here, but it is clear from the above that our boundary conditions are satisfied by any solution which can be obtained by acting with a supersymmetry transformation on a member of the above class of $\psi_\mu=0$ solutions.   This class of solutions is sufficiently broad to define an interesting phase space.

\medskip
Since the symplectic flux through $\I$ vanishes, the symplectic
structure is finite, and the supersymmetry variations leave the
boundary conditions invariant, we know from the general analysis of
the previous section that a conserved supercharge will exist. We now
follow the procedure described in the previous section to determine
what those charges are. The Noether current 3-form ${\bf J}$ is
given by \ben {\bf J}_\eta(\Phi) = \bt(\Phi, \b_\eta \Phi) -
\B_\eta(\Phi) = 2\bar \eta \Gamma {\bf E} \, , \een where $(^* {\bf
E})_\mu$ are the equations of motion of the spin 3/2-field.
Consequently, the Noether current vanshes identically on shell.
Thus, the Noether charge vanishes on shell, too, $\Q_\eta=0$. Using
the defining relation~\eqref{Adef} for $\d \A_\eta$, one finds
 \ben\label{N} (\d  \A_\eta)_{\mu\nu\sigma} =
 \delta[\epsilon_{\mu\nu\sigma\alpha} (\nabla_\beta {\bf
 N}^{\alpha\beta})], \quad {\bf N}_{\alpha\beta} = \bar \eta
 \gamma_{[\alpha} \gamma_{\beta} \gamma_{\rho]}
 \psi^\rho \, \,   ,
 \een for any on-shell variation of the fields, and so we can read
off \ben \A_\eta = \delta (^* {\bf N}) \, . \een In particular, we
explicitly see that the consistency condition holds. Because the
Noether charge vanishes on shell, according to eq.~\eqref{Hdef} the
conserved charge associated with $\eta$ is now given by 
\ben \H_\eta
= \int_{\partial \Sigma}{} ^* {\bf N} 
= 2\int_{\partial \Sigma}
(\bar \eta \gamma_{[\nu} \gamma_\sigma \gamma_{\rho]} \psi^\rho) 
u^{\nu\sigma} \,
\d^2 S \, . 
\een 
Here, $u^{\mu\nu}$ is the binormal to $\partial \Sigma$ (normalized
with respect to the physical metric), and $\d^2 S$ the induced volume
element.
The fall-off properties imposed upon the
fields as part of the boundary conditions guarantee that this
integral is finite. By eq.~\eqref{offshell}, the off-shell
Hamiltonian is obtained by adding to the above expression the
constraints. Since $\J_\eta = -2\bar \eta \Gamma \E$, it follows
that those are given by $\C = 2\Gamma \E$, so the off-shell spinor
charge is 
\ben 
\H_\eta = 2 \int_\Sigma \bar \eta \left[
\gamma_{[\mu} \nabla_\nu \psi_{\sigma]} -\frac{1}{4} T_{\tau[\mu\nu}
\gamma^\tau \psi_{\sigma]} \right] \epsilon^{\mu\nu\sigma \alpha}
u_\alpha \, \d^3 S + 2 \int_{\partial \Sigma} (\bar \eta
\gamma_{[\nu} \gamma_\sigma \gamma_{\rho]} \psi^\rho)
u^{\nu\sigma}\, \d^2 S \, , 
\een 
where $T_\mu$ is the torsion 1-form
of $D_\mu$, where $u^\mu$ is the unit normal to $\Sigma$, where
$u^{\mu\nu}$ the normalized bi-normal to $\partial \Sigma$, and
where $\d^3 S$ and $\d^2 S$ are the respective integration elements
induced from the physical metric. This agrees with the expression
found by~\cite{Henneaux1985} using a different method\footnote{Note,
however, that the boundary
  conditions on $\psi_\mu$ imposed in~\cite{Henneaux1985} are stronger
than ours.}.

\section{Extended supergravity}

We would now like to extend the above analysis to extended
supergravities~\cite{Das1977,Freedman1977,DeWit1981a,DeWit1981b,Cremmer1976}.
In the extended ${\mathcal N} = 4$ supergravity
theory, the fields are a metric, four spin 3/2 fields
$\psi^{(a)}_\mu$,
4 Majorana spin 1/2 fields $\chi^{(a)}$, a real scalar field $A$ and a
real pseudoscalar field $B$, and an $SO(4)$ gauge connection
$A^{(a)(b)}_\mu$, where $a,b=1,\dots,4$.
The Lagrangian consists of (\ref{min})
plus a complicated matter Lagrangian, which we
shall not write down. The full supersymmetry transformations are
likewise very complicated. Fortunately, for the present analysis, only the
leading parts will be important, and so we will not need to
write out the detailed Lagrangian, and transformation rules.
%Those are given by
%\bena
%\b_\eta g_{\mu\nu} &=& -2i \bar \eta^{(a)} \gamma_{(\mu} \psi^{(a)}_{\nu)}\\
%\b_\eta \psi^{(a)}_\mu &=& \left(D_\mu + \frac{1}{2} \gamma_\mu\right)\eta^{(a)}
%- \frac{i}{2 \cdot 4} [\gamma^\lambda, \gamma^\tau]
%F_{\lambda\tau}^{(a)(b)} \gamma_\mu \eta^{(b)} + \dots\\
%\b_\eta \chi^{(a)} &=& -\frac{1}{\sqrt{2}} [
%i \gamma^\mu \nabla_\mu ( A + i\Gamma B) +
%\ell(A- i\Gamma B) ] \eta^{(a)} + \dots \\
%&-& \frac{1}{2 \cdot 4 \sqrt{2}} \epsilon^{(a)(b)(c)(d)}
%[\gamma^\mu, \gamma^\nu] \eta^{(b)} F^{(c)(d)}_{\mu\nu} + \dots\\
%\b_\eta A &=& \frac{1}{\sqrt{2}} \bar \eta^{(a)} \chi^{(a)} + \dots \\
%\b_\eta B &=& \frac{i}{\sqrt{2}} \bar \eta^{(a)} \Gamma \chi^{(a)} +
%\dots\\
%\b_\eta A^{(a)(b)}_\mu &=& \frac{i}{\sqrt{2}} \epsilon^{(a)(b)(c)(d)}
%\bar \eta^{(c)} \gamma_\mu \chi^{(d)} - 2 \bar\eta^{[(a)} \psi^{(b)]}_\mu
%\eena
%where omitted terms are at least quadratic in the fields
%$A,B, A_\mu^{(a)(b)},\psi^{(a)}_\mu$, and will be of higher order in
%$\Omega$.

We would like to pick a set of boundary conditions for the above
fields such that the symplectic form is finite, such that the
symplectic current through $\I$ vanishes, and such that the boundary
conditions are invariant under ${\mathcal N}=1$ supersymmetry.
The essence of this problem can be studied in a scalar reduction
of the ${\mathcal N}=4$ theory, and we will from now on limit ourself
to that reduction. In the scalar reduction~\cite{Das1977}, only
the fields $\chi = d^{(a)} \chi^{(a)}$, $A, B$, $g_{\mu\nu}$, and
$\psi_\mu = d^{(a)} \psi^{(a)}_\mu$ are kept, where $d^{(a)}$ is a
given direction in field space in which
we want to preserve the ${\mathcal N}=1$ supersymmetry, and
all other fields are set to zero. The precise Lagrangian, and
supersymmetry transformation rules of this reduced theory are
as follows, see section~4
of~\cite{Das1977}:
\ben\label{lagr}
\LL = \LL_{min} + \LL_{kin} + \LL_{int} \, .
\een
Here, $\LL_{min}$ is identical to the Lagrangian of minimal
supergravity, see eq.~\eqref{min}, without the cosmological
constant term. $\LL_{kin}$ contains the kinetic terms for the matter
fields,
\bena\label{kin}
\LL_{kin} &=& \beps \Bigg\{ -i \bar \chi \gamma^\mu \nabla_\mu \chi
-\frac{1}{(1-u)^2}[(\nabla_\mu A) \nabla^\mu A + (\nabla_\mu B)
\nabla^\mu B] \non\\
&&+\frac{i \sqrt{2}}{1-u} \bar \psi_\nu \nabla_\mu(A + i \Gamma B) \gamma^\nu \gamma^\mu\chi \\
&&+ \frac{i}{1-u}
\epsilon^{\lambda \mu \nu \rho} \bar \psi_\lambda \gamma_\mu \psi_\nu
(A \nabla_\rho B - B \nabla_\rho A) \non\\
&&+ \frac{5}{2(1-u)} \bar \chi \Gamma \gamma^\rho \chi \,
(A \nabla_\rho B - B \nabla_\rho A) \Bigg\} \, ,\non
\eena
where $u=A^2+B^2$.
The interaction Lagrangian is
\bena\label{int}
\LL_{int} &=& \frac{3-u}{1-u} \beps +
\frac{i\sqrt{2}}{(1-u)^{1/2}}
\bar \psi_\mu \gamma^\mu (A+i\Gamma B) \chi \, \beps\non\\
&&+\frac{i}{2(1-u)^{1/2}} \bar \psi^\mu [\gamma_\mu,\gamma_\nu]
\psi^\nu \, \beps+ \dots\, .
\eena
where the dots stand for terms that are dropping off fast at infinity
and can be ignored for our purposes. The first term in $\LL_{int}$
provides a negative cosmological constant $\Lambda=-3$,
and gives a mass $m^2_A=-2=m_B^2$
to linear fluctuations of the scalar fields $A$ and $B$ about $0$.

The terms in the supersymmetry transformation law for
$(A,B,\chi)$ that will be relevant for us are given by
\bena
\b_\eta \chi &=& -\frac{1}{\sqrt{2}} [
%i
\gamma^\mu \nabla_\mu ( A + i\Gamma B) +
(A- i\Gamma B) ] \eta + \dots \\
\b_\eta A &=& \frac{i}{\sqrt{2}} \bar \eta \chi + \dots \\
\b_\eta B &=& \frac{-1}{\sqrt{2}} \bar \eta \Gamma \chi + \dots \, .
\eena
The dots stand for terms that fall off faster at infinity and are not
relevant for the considerations below.
%For the vector fields, defining $A^{(a)}_\mu = d^{(b)} \epsilon^{(a)
%(b)(c)(d)} A^{(c)(d)}_\mu$, so that $(a)=1,2,3$ is now an index
%perpendiclar to $d^{(a)}$ the transformations become
%\bena
%\b_\eta \chi^{(a)} &=& \frac{1}{\sqrt{2} 4} [\gamma^\mu,\gamma^\nu]
%F^{(a)}_{\mu\nu \eta} \,   \\
%\b_\eta A_\mu^{(a)} &=& \frac{i}{\sqrt{2}} \bar \eta \gamma_\mu
%\chi^{(a)} \, \quad (a) = 1,2,3 \, .
%\eena
For the spin-3/2 field and metric, the supersymmetry variations
take the form
\bena\label{psivar}
\b_\eta g_{\mu\nu} &=& - 2i \bar \eta \gamma_{(\mu} \psi_{\nu)} \\
%\b_\eta \omega_{\mu \, ij} &=& S_{\mu \, ij} -
%S_{\tau [i}{}^\tau e_{j] \mu}\\
\b_\eta \psi_\mu &=& \left(D_\mu +
\frac{1}{2(1-u)^{1/2}} \gamma_\mu + \frac{1}{2(1-u)}
[A \nabla_\mu B - B \nabla_\mu A] \Gamma \right) \eta
+\dots\, ,
\eena
where again, dots stands for terms of higher order in $\Omega$
that can be neglected for our purposes, and where we have
set $u = A^2 + B^2$. We will now discuss in detail the boundary
conditions for the scalar multiplet.

\subsection{The scalar multiplet}

From the Lagrangian given above, eq.~\eqref{kin}, one derives the
following expression for the symplectic current containing
contributions from the fields $A,B,\chi$:
\bena
^* \bomega_\rho &=&  -2i \delta_1 \bar \chi \gamma_\rho \delta_2 \chi
-2 \delta_1 A \nabla_\rho \delta_2 A -2 \delta_1 B \nabla_\rho \delta_2
B\non\\
&& +i\sqrt{2} \delta_2 \bar \psi_\nu \delta_1 A \gamma_\rho
\gamma^\nu \chi \frac{1}{1-u} + i\sqrt{2} \bar \psi_\nu
\gamma_\rho \gamma^\nu \delta_1 A \delta_2 \left( \chi \frac{1}{1-u} \right)\non\\
&& + (A \leftrightarrow i\Gamma B)\non\\
&&+ i\delta_2 \left( \epsilon^{\lambda\mu\nu}{}_\rho
\bar \psi_\lambda \gamma_\mu \psi_\nu \frac{1}{1-u}
\right)[A\delta_1 B - B \delta_1 A] \non\\
&&+ \delta_2 \left( \bar \chi \Gamma \gamma_\rho \chi \frac{5}{2(1-u)} \right)
[A\delta_1 B - B\delta_1 A] \non\\
&& - (1 \leftrightarrow 2) \, .
\eena
Linear fluctuations of the scalar fields $A,B$
have a mass of $m^2_A = -2 = m^2_B$. An analysis of possible boundary
conditions for linearized fields on exact AdS spacetime with that mass
indicate the following fall-off behavior at the
linearized level~\cite{BF,IW}:
\bena
\label{481} \tilde A \equiv \Omega^{-1} A &=& \alpha_A + \Omega \beta_A + \dots\\
\label{482}  \tilde B \equiv \Omega^{-1} B &=& \alpha_B + \Omega \beta_B + \dots \,.
\eena
 We will consider only boundary conditions consistent with (\ref{481}) and (\ref{482}). The
  contributions to the symplectic current 3-form
from the scalar fields are  then finite at $\I$, as may be seen from
the expression \ben \bomega_{\mu\nu\sigma} \restriction \I =
[\delta_1 \alpha_A \delta_2 \beta_A + \delta_1 \alpha_B \delta_2
\beta_B- (1 \leftrightarrow 2) ] \, \tilde n^\rho \tilde
\epsilon_{\rho\mu\nu\sigma} \een for the contribution symplectic
current involving the fields $A$ and $B$. The corresponding flux
through scri is \ben F_{spin-0} = 2\int_{\I} [\delta_1 \alpha_A
\delta_2 \beta_A + \delta_1 \alpha_B \delta_2 \beta_B - (1
\leftrightarrow 2)] \, \d^3 S \, , \een where $\d^3 S$ is the
integration element on $\I$ induced by the unphysical metric $\tilde
g_{\mu\nu}$. The flux $F_{spin-0}$ will not vanish unless there is a
relationship between $(\alpha_A, \alpha_B, \beta_A, \beta_B)$. One
possible relationship that will guarantee the vanishing of the flux
is to impose
 \ben \label{abq} \alpha_A = q \alpha_B,
 \quad \beta_B = -q \beta_A \, ,
 \een where $q \in \mr \cup \{\infty\}$. Each choice
of $q$ corresponds to a particular boundary condition. This defines
a Lagrange submanifold in the space $(\alpha_A, \alpha_B, \beta_A,
\beta_B)$ with the symplectic form given above. However, other
Lagrange submanifolds may be chosen as well.

\medskip

Linear perturbations of the spinor field $\chi$ off of exact AdS
spacetime may be used to identify a class of reasonable boundary
conditions. Such an analysis indicates that~\cite{BF}
\ben
\tilde \chi = \Omega^{-3/2} \chi
\een
should be finite (for example smooth) at $\I$. A simple way of seeing this
is to note that the linearized field $\chi$ satisfies a massless
Dirac equation on AdS-spacetime. The massless Dirac equation in
4 dimensions is conformally invariant, with conformal weight
$3/2$. This implies that if $\chi$ is a smooth solution to the
massless Dirac equation in AdS spacetime, then $\tilde \chi$ is a
smooth solution on the conformal compactification.
The contribution to symplectic current from the spinor field $\chi$ at $\I$
is then expressible as
\ben
\bomega_{\mu\nu\sigma}
= i
[\overline{\delta_1 \tilde \chi} \tilde \gamma^\alpha  \delta_2 \tilde
\chi - (1 \leftrightarrow 2)] \tilde \epsilon_{\alpha\mu\nu\sigma} \, .
\een
This will be finite at $\I$ under the above choice of
boundary condition that  $\tilde \chi = \Omega^{-3/2} \chi$ be smooth
at $\I$. The symplectic flux is given by
\ben
F_{spin-1/2} = \int_\I [\overline{\delta_1 \chi_+} \Gamma
\delta_2 \chi_- - \overline{\delta_1 \tilde \chi_-} \Gamma \delta_2
\chi_+ -(1 \leftrightarrow 2)] \, \d^3 S \, ,
\een
where we have defined positive and negative chiral projections by
\ben
\chi_+ = \tilde P_+ \tilde \chi, \quad
\chi_- = i \tilde P_+ \Gamma \tilde \chi = i \Gamma \tilde P_- \tilde \chi\, .
\een
The spin-1/2 symplectic
flux will only vanish if there is a relationship between these
projections, for example
\ben
\chi_- = q \chi_+ \, ,
\een
but, again, more general boundary conditions described
by an arbitrary Lagrange submanifold in the space $(\chi_+,
\chi_-)$ would also ensure the vanishing of $F_{spin-1/2}$.
As above, each $q$ defines a separate boundary condition.
There is in particular
no reason at this stage why the $q$ in this equation must be
equal to the $q$ in
eq.~(\ref{abq}).  But we will now see
that, if they are equal, then the boundary conditions preserve ${\mathcal N}=1$
supersymmetry. For this, we need to find how the supersymmetry
transformations act on the boundary values of the fields in the
chiral multiplet,
$(\alpha_A, \alpha_B, \beta_A, \beta_B, \chi_+, \chi_-)$.

To this end, we now choose an asymptotic Killing spinor $\eta=
\Omega^{-1/2}\tilde\eta$, and define the quantities
$\eta_+ = \tilde P_+ \tilde \eta$, and $\eta_- = \frac{1}{2} \tilde
P_- \tilde \eta$, both of which are finite at $\I$ by the arguments
given above around eq.~\eqref{gauge}. We insert this
into formulae for the supersymmetry transformations. After some
algebra, using the field equations for the fermion field, we find,
\bena\label{bsusy}
\b_\eta \alpha_A &=& \frac{i}{\sqrt{2}} \bar \eta_+ \chi_-\\
\b_\eta \alpha_B &=& \frac{i}{\sqrt{2}} \bar \eta_+ \chi_+\\
\b_\eta \beta_B &=&  \frac{-1}{2\sqrt{2}} \bar \eta_- \Gamma \chi_-
+ \frac{1}{\sqrt{2}}
\bar \eta_+ \Gamma \tilde h^{\mu\nu} \tilde
\gamma_\mu \tilde \nabla_\nu \chi_-
\\
\b_\eta \beta_A &=& \frac{+1}{2\sqrt{2}} \bar\eta_- \Gamma \chi_+
- \frac{1}{\sqrt{2}}  \bar \eta_+
\Gamma \tilde h^{\mu\nu} \tilde \gamma_{\nu} \tilde \nabla_\mu \chi_-\\
\b_\eta \chi_- &=& \frac{-1}{\sqrt{2}} \left[
+ \alpha_A \eta_+ + i\Gamma \beta_B \eta_+ -  \tilde h^{\mu\nu} \tilde \gamma_\mu
\tilde \nabla_\nu \alpha_A \eta_+
\right] \\
\b_\eta \chi_+ &=& \frac{-1}{\sqrt{2}} \left[ + \alpha_B \eta_-
-i\Gamma \beta_A \eta_+ -  \tilde h^{\mu\nu} \tilde \gamma_\nu
\tilde \nabla_\mu \alpha_B \eta_+ \right] .\eena Here, $\tilde
h_{\mu\nu}$ is the metric of the ESU, meaning that all derivatives
in the above formulae are tangent to $\I$, and hence well-defined on
the boundary fields. It now easy to see that the above boundary
conditions are preserved by supersymmetry. For example $\alpha_A =
q\alpha_B$ and $\beta_B = - q\beta_A$ immediately imply that
$\b_\eta \chi_- = q \b_\eta \chi_+$, which is just the desired
boundary condition for the spin-1/2 field. The bosonic part of the
asymptotic symmetry group associated with these boundary conditions
is isomorphic to $O(3,2)$.

\medskip
\noindent
We will now show that there exist further boundary conditions
that preserve a subset
of the ${\mathcal N} = 1$ supersymmetry transformations associated
with a subgroup $O(2,1) \times \mr^3$ of $O(3,2)$.
For this,
we first note that the supersymmetry transformations
mix only the boundary fields $(\alpha_A, \beta_B, \chi_-)$
and $(\alpha_B, \beta_A, \chi_+)$, but do not mix fields from
one set with fields from the other set. This suggests to
combine each set into a boundary superfield, and to try and write the
supersymmetry transformations on the boundary fields in terms of a
derivative operator on superspace. We will now show that this can
indeed be done.

For this, we now consider a 2-dimensional subspace of the
4-dimensional space of the asymptotic Killing
spinors $\eta$ that parametrize our supersymmetry
transformations. Recall that in exact AdS-space, there are four real
linear independent solutions to the Killing spinor
equation~\eqref{bgspinor}. The corresponding conformally rescaled
spinors $\tilde \eta = \Omega^{1/2} \eta$ satisfy the conformally
related equivalent equation~\eqref{gauge0}. To single out the
specific 2-dimensional
subspace of Killing spinors on AdS-spacetime, we consider
the conformally rescaled Killing-spinor equation~\eqref{gauge0} for
the metric $\d \tilde s_0^2$, related to the exact AdS-spacetime
$\d s^2_0$ via the conformal factor $\Omega=z$ defined by writing the AdS
metric in Poincare coordinates,
\ben\label{poincare}
\d s^2_0 = \frac{1}{z^2}(-\d t^2 + \d x^2 + \d y^2 + \d z^2) = \Omega^2
\d \tilde s^2_0 \, , \quad
z \ge 0 \, .
\een
In this description, AdS-spacetime is mapped to
the half space $\mr^3 \times \mr_+$ of
4-dimensional Minkowski spacetime. The conformal Killing spinor
equation~\eqref{gauge0} in this description (with $\Omega = z$) has as obvious
solutions the two linearly independent constant spinors
$\tilde \varepsilon=\tilde \eta$ in
4-dimensional Minkowski spacetime satisfying $\tilde P_- \tilde \varepsilon =
0$, and $\tilde P_+ \tilde \varepsilon = \tilde \varepsilon$
where we note that $\tilde P_\pm$ is now simply the projector
$(1\pm\gamma_z)/2$, with $\gamma_z$ the flat Minkowski space
gamma-matrix. The two real linear independent Killing spinors
$\varepsilon = \Omega^{-1/2} \tilde
\varepsilon$
thus obtained on AdS-spacetime
generate the 3 translations $\partial_t, \partial_x, \partial_y$ on
the conformal boundary defined by $z=0$ under the supersymmetry
algebra, i.e., $[s_{\varepsilon_1}, s_{\varepsilon_2}] = \pounds_X$,
where $X=\bar \varepsilon_1 \gamma^\mu \varepsilon_2 \partial_\mu$
is a translation on $\mr^3$, see also~\cite{LP1,LP2,Muck1999}.

Consider
%now
the boundary fields $(\alpha_A, \alpha_B, \beta_A,
\beta_B, \chi_+, \chi_-)$, viewed now as fields on $\mr^3$ via the
diffeomorphism $(t,r,\theta,\phi) \to (t,x,y,z)$,
which maps the conformal infinity $\I=\mr \times S^2$
(minus a generator) to $\mr^3$. They may be obtained from the boundary
fields defined above on $\I$ via the conformal change
\bena\label{ktrafo}
&&\chi_\pm \to k^{3/2} \chi_\pm\\
&&\alpha \to k \alpha \\
&&\beta \to k^2 \beta \, ,
\eena
followed by the diffeomorphism, where $k$ is the ratio of the
conformal factors in eq.~\eqref{poincare} and~\eqref{global},
and $\beta$ means either $\beta_A$ or $\beta_B$, etc.
For any one of the 2 real linear
independent Killing spinors $\varepsilon$ just described,
we have
\bena\label{bsusy1}
\b_\varepsilon \alpha_A &=& \frac{i}{\sqrt{2}} \overline{\tilde\varepsilon}
\chi_-\\
\b_\varepsilon \alpha_B &=& \frac{i}{\sqrt{2}} \overline{\tilde\varepsilon}  \chi_+\\
\b_\varepsilon \beta_B &=&  + \frac{1}{\sqrt{2}}
\overline{\tilde\varepsilon} \Gamma \tilde h^{\mu\nu} \tilde
\gamma_\mu \tilde \nabla_\nu \chi_-
\\
\b_\varepsilon \beta_A &=& - \frac{1}{\sqrt{2}}  \overline{\tilde\varepsilon}
\Gamma \tilde h^{\mu\nu} \tilde \gamma_{\nu} \tilde \nabla_\mu \chi_+\\
\b_\varepsilon \chi_- &=& \frac{-1}{\sqrt{2}} \left[
+ i\Gamma \beta_B  -  \tilde h^{\mu\nu} \tilde \gamma_\mu
\tilde \nabla_\nu \alpha_A
\right] \tilde \varepsilon \\
\b_\varepsilon \chi_+ &=& \frac{-1}{\sqrt{2}} \left[ -i\Gamma
\beta_A  -  \tilde h^{\mu\nu} \tilde \gamma_\nu \tilde \nabla_\mu
\alpha_B \right] \tilde \varepsilon \, . \label{bsusy2}
 \eena
Note that in this expression $\tilde h_{\mu\nu}$ is now the metric
on the conformal boundary $z=0$ of $\mr^3$ in~\eqref{poincare},
i.e., the flat, 3-dimensional Minkowski metric. These
transformations may be written in terms of standard superfields on
3-dimensional Minkowski spacetime $\mr^3$. For this, introduce 3
Pauli-matrices $\sigma_j$, $j=t,x,y$ intrinsic to  $\mr^3$, identify
$\tilde \varepsilon, \tilde \chi_\pm$ with real 2-component spinors
on $\mr^3$ associated with the group $SO(2,1)$ of isometries of the
boundary $\mr^3$, and introduce real Grassmann valued spinor field
$\theta^\alpha$. Define two real boundary superfields on $\mr^3$ by
 \bena
 \label{BSF1}
 \Phi_+ &=& \alpha_A + \bar \theta \chi_+  + \frac{1}{2} \theta^2 \beta_B , \\
 \label{BSF2} \Phi_- &=& \alpha_B + \bar \theta \chi - \frac{1}{2}
 \theta^2 \beta_A .
 \eena Define a superspace derivative operator as
usual by \ben\label{superspace} {\mathcal D}_\alpha =
i\frac{\partial}{\partial \theta^\alpha} +\sigma^{j}{}_{\alpha
\delta} \theta^\delta \partial_j  . \een Then the SUSY
transformations of the boundary fields can be written as
\ben\label{superd} \b_\varepsilon \Phi_\pm = \frac{1}{\sqrt{2}}
\tilde \varepsilon^\alpha {\mathcal \D}_\alpha \Phi_\pm \, . \een
The above choice of the boundary conditions is simply \ben \Phi_+ =
q \Phi_- \, , \een and this is now manifestly invariant under SUSY
transformations with respect to the two linear independent spinors
$\varepsilon$, since these are given in terms of a superderivative.
We note that this formalism is in direct parallel to the
three-dimensional superspace formalism used to study an AdS/CFT dual
of our discussion in \cite{LP1,Muck1999}.

The most general class of boundary conditions preserving the
supersymmetry transformations generated by the two Killing spinors
$\varepsilon$ can now be inferred as
follows. The first condition on any boundary condition was that the
symplectic flux through $\I$ vanishes. The combined spin-0 and spin-1/2
contributions can be written in terms of the superfields as
\ben
F_{spin-0} + F_{spin-1/2}
= 2\int_\I  [\delta_1 \Phi_+ \delta_2 \Phi_-
- \delta_1 \Phi_- \delta_2 \Phi_+] \, \d^2 \theta \d^3 S \, .
\een
It follows from this superspace
expression for the symplectic flux that it will vanish if we impose as a
boundary condition
\ben\label{sls}
\Phi_- = \frac{\delta {\mathcal W}(\Phi_+)}{\delta \Phi_+} \, ,
\een
for some functional ${\mathcal W}$ of $\Phi_+$. Indeed, we then have
\ben
F_{spin-0} + F_{spin-1/2} =
2\int  \frac{\delta^2 {\mathcal W}}{\delta \Phi_+ (x_1) \delta
  \Phi_+(x_2)} [\delta_1 \Phi_+(x_1) \delta_2 \Phi_+(x_2)
- \delta_1 \Phi_+(x_2) \delta_2 \Phi_+(x_2)] = 0
\, .
\een
If in addition, the functional dependence of $\mathcal W$ on
the boundary fields is of the form
\ben\label{Wdef}
{\mathcal W}(\Phi_+) = \int_{\I} {\mathcal F}(\Phi_+) \,
\d^2 \theta \d^3 S \, ,
\een
(i.e., $\mathcal W$ depends on the boundary fields only through the superfield)
then the boundary condition is
also invariant under the operation $\mathcal D$, i.e., under
supersymmetry.  Thus, we have found that
the above ``supersymmetric Lagrange submanifold''
condition gives a general form for
boundary conditions perserving ${\mathcal N}=1$ supersymmetry.
When expressed in terms of the
boundary fields, our supersymmetric boundary conditions are as follows.

\paragraph{SUSY boundary conditions for scalar multiplet:}
Let $\Omega$ denote the conformal factor in~\eqref{poincare}, so that
the conformal boundary is mapped to $\mr^3$.
Define the boundary fields $\alpha_A, \alpha_B, \beta_A, \beta_B,
\chi_+, \chi_-$ by
\bena\label{ABas}
A &=& \Omega \alpha_A + \Omega^2 \beta_A + \dots\\
B &=& \Omega \alpha_B + \Omega^2 \beta_B + \dots\\
\chi_+ &=& \Omega^{-3/2} \tilde P_+ \chi + \dots\\
\chi_- &=& i \Omega^{-3/2} \Gamma \tilde P_- \chi + \dots \, .
\eena
Let $\varepsilon$ be the two Killing spinors that approach the
constant spinors on the boundary $\mr^3$. Then, for any
smooth function $\F(z)$ in one variable, the boundary conditions
\bena\label{bndycond}
\alpha_A &=& {\mathcal F}'(\alpha_B) \, , \\
\beta_B &=& -\beta_A {\mathcal F}''(\alpha_B) + \frac{1}{2} {\mathcal
  F}'''(\alpha_B) \bar \chi_+ \Gamma \chi_+ \, , \\
\chi_- &=& {\mathcal F}''(\alpha_B) \chi_+\, ,
\eena
leave ${\mathcal N}=1$ supersymmetries generated by the 2 real linear
independent Killing spinors $\varepsilon$ unbroken. They are the component
versions of the supersymmetric Lagrange submanifold  condition $\Phi_-=
\delta {\mathcal W}/\delta \Phi_+$, see eq.~\eqref{Wdef}.

\medskip
\noindent
It can be seen that the
asymptotic symmetry algebra associated
with these boundary conditions is the superalgebra
associated with the bosonic asymptotic symmetry group $O(2,1) \times \mr^3$
that induces the isometries of the
boundary metric $-\d t^2 + \d x^2 + \d y^2$, see~\eqref{poincare}.
Thus, in particular,  conserved bosonic and fermionic generators
of this algebra exist on phase space, and can be constructed according
to the algorithm described in sec.~\ref{defs}.
If we make the special
choice $\F(z) = \frac{1}{2} q z^2$, so that
${\mathcal W} = \int \frac{1}{2} q \Phi_+^2 \, \d^2
\theta \d^3 S$, then we obtain the set of boundary conditions invariant
under the larger asymptotic symmetry superalgebra with bosonic
part $O(3,2)$ generated by
supersymmetry transformation with arbitrary Killing spinor parameter
$\eta$ that was already described above. In particular, for $q=0$ or
$q=\infty$, we recover the boundary condition proposed by
Breitenlohner and Freedman~\cite{BF} (see also~\cite{Hawking1983})
as a special case of the more general
boundary conditions shown above to preserve ${\mathcal N}= 1$
supersymmetry.

The above boundary conditions can equivalently be formulated in terms
of the boundary fields associated with the conformal completion
$\I=S^2\times\mr$. The corresponding conditions are obtained
from~\eqref{bndycond} by simply making the transformations~\eqref{ktrafo}.
One may also implement the full supersymmetry transformations~\eqref{bsusy}
associated with an arbitrary Killing spinor field $\eta$ by a
superspace derivative operator, which would now be of the form
$s_\eta = \eta_+ \partial/\partial \theta +
(\eta_+ \sigma^i \theta) \partial_i + 2\eta_- \theta$.
Comparing to the superspace derivative~\eqref{superd}, there is now
an additional term $2\theta \eta_-$, and this explains why a general
function $\F$ other than $\F(z)=\frac{1}{2} q z^2$ no longer gives
boundary conditions that are invariant under $s_\eta$ for arbitrary
asymptotic Killing spinors $\eta$.

\subsection{The gravity multiplet}

The boundary conditions for the gravity multiplet $(\psi_\mu, g_{\mu\nu})$
are very similar in nature to those in the minimal
supergravity theory discussed above, and indeed are identical
for the metric. There does, however,
exist a subtle difference for the boundary condition of the
spin-3/2 field having to do with the backreaction of the
scalar fields, and this arises as follows. Above, we discussed the fact
that the vanishing of the symplectic flux will hold if
we have eq.~\eqref{vani}, and this suggested to take
$\psi_\mu = O(\Omega^{3/2})$ as the boundary condition for the
spin-3/2 field in the minimal supergravity theory.
When we checked whether this boundary condition
is also preserved by supersymmetry, we found that the
supersymmetry variation $\b_\eta \psi_\mu$ was given by
eq.~\eqref{var}. A priori, this did not satisfy the desired
boundary conditions, which would require a faster fall-off
in~\eqref{var} by one power. However, it then followed from the
field equation that, in fact $\Omega^{-1} \tilde K_{\mu\nu} = 0$
at $\I$, so
%the faster fall of indeed holds.
the predicted faster fall-off condition does indeed hold.

In the extended supergravity theory, the same reasoning does not go
through as it stands, because the field equations do not
imply any more that $\Omega^{-1} \tilde K_{\mu\nu} = 0$. The reason
for this is that, while the stress tensor vanished at $\I$ in the
minimal supergravity, we now have a non-vanishing contribution
arising from the scalar field. Instead, using the same expansion
techniques as above around~\eqref{recursion},
together with the boundary conditions
for the scalar and spin-1/2 fields, we now infer that
\ben\label{k1}
\Omega^{-1} \tilde K_{\alpha\beta}
= -\frac{1}{4}(\alpha_A^2 + \alpha_B^2) \tilde h_{\alpha\beta} \, .
\een
where $\tilde h_{\mu\nu}$ is the induced metric at $\I$, i.e., the
metric of the ESU. This does not vanish. Consequently, using
this in~\eqref{psivar}, we now get for the susy variation of the
spin-3/2 field in the reduced ${\mathcal N}=4$ supergravity
\bena\label{psivar'}
\b_\eta \psi_\mu &=& \left( D_\mu + \left[ \frac{1}{2} + \frac{1}{4} u
    + \dots \right] \gamma_\mu \right) \eta \non\\
&&+
\frac{1}{2} (1+u+\dots)
(A \nabla_\mu B - B \nabla_\mu A) \Gamma \eta + \dots
\non\\
&=& -\Omega^{1/2} \frac{1}{8}(\alpha_A^2 + \alpha_B^2) \tilde
n_{\mu} \tilde P_+ \tilde U + O(\Omega^{3/2}) \eena where we have
again used the asymptotic form of the fields $A,B$, and the the
metric, see eq.~\eqref{asymptg} below. Thus, unlike in the minimal
supergravity theory, the supersymmetry variation of the spin-3/2
field is only of order $O(\Omega^{1/2})$ even after taking into
account the field equations. Thus, if we are to choose supersymmetry
invariant boundary conditions, we must impose

\paragraph{Boundary conditions for $\psi_\mu$:} We have
$\psi_\mu = \Omega^{1/2} \tilde n_\mu \tilde \psi_+ +
O(\Omega^{3/2})$, where $\tilde \psi_+$ is smooth at $\I$
and satisfies $\tilde P_- \tilde \psi_+ = 0$.

\medskip
\noindent

The point is now that, since the slow fall off piece in
$\psi_\mu$ is proportional to $\tilde n_\mu$, the
condition~\eqref{vani} for the vanishing of the symplectic flux
still holds. The supersymmetry variation of the metric also still
satisfies our boundary conditions even with the above weaker boundary
condition for the spin-3/2 field. This may be seen e.g. by noting
that $(\b_\eta g_{\mu\nu}, \b_\eta \psi_\mu)$ satisfy the linearized
equations of motion. The linearized
version of our above expansion techniques then show that $\b_{\eta}
g_{\mu\nu}$ is finite at $\I$, and thus satisfies the linearized
boundary conditions for the metric. Alternatively, we may directly calculate using the
boundary conditions for the spin-3/2 field that
\ben
\b_\eta g_{\mu\nu} = \Omega^{-1} \tilde n_{(\mu} \tilde X_{\nu)} +
\dots \, ,
\tilde X_\mu = \overline{\tilde \psi_+} \tilde \gamma_\mu \tilde \eta
\, .
\een
Noting that $\tilde n^\mu \tilde X_\mu = 0$ at scri, and putting
$X^\mu = \frac{1}{2} \Omega^2 \tilde g^{\mu\nu} \tilde X_\nu$, one
can then easily see that
\ben
\Omega^{-1} \tilde n_{(\mu} \tilde X_{\nu)}  = \pounds_X g_{\mu\nu} +
O(\Omega^0) \, .
\een
Thus, up to an
 infinitesimal diffeo  the susy variation of the
metric is of order $O(\Omega^0)$, which is the desired
boundary condition for the metric.
The additional term $\pounds_X g_{\mu\nu}$ can be dealt with in either of two ways:  1) Note that $\pounds_X$ changes $g_{\mu \nu}$ only at order $O(\Omega^{-1})$.  Thus, one may slightly weaken our original boundary condition allowing a departure from ``Gaussian normal gauge'' at this order.  This makes the recursion describing the expansion in powers of $\Omega$ somewhat more complicated, but does not significantly change the results.  2) One may take the physical symmetry to be not $s_\eta$, but $s_\eta - \pounds_X$.  One then readily checks that the
$s_\eta - \pounds_X$ preserve the boundary conditions on all fields, and that they satisfy the same algebra as the original $s_\eta$.

\section{Expressions for the bosonic and
fermionic charges in ${\mathcal N}=4$
extended supergravity}

\subsection{Bosonic charges}
Above, we have formulated boundary conditions for the fields
$A,B,\chi$, and $g_{\mu\nu}, \psi_\mu$, in terms of a
free function $\mathcal F$. The group of (bosonic) symmetries
leaving these invariant was $O(3,2)$ for the special choice
${\mathcal F}(z) = \frac{1}{2} q z^2$,
and it was $O(2,1) \times \mr^3$ for general $\mathcal F$.
These boundary conditions were chosen in such a way that the
symplectic flux through $\I$ vanishes. By the general analysis of
Sec.~\ref{defs}, we therefore know that Hamiltonian generators
for these symmetries exist. We will now find the explicit
expressions for those
conserved bosonic charges, starting with the case of generic $\F(z)$.

The corresponding Lie algebra of $O(2,1) \times \mr^3$ of asymptotic
bosonic symmetries acts by vector fields $X$ on $M$ that are
smooth and tangent to $\I$, and are Killing fields of the induced metric on
$\I$. These vector fields correspond to
translational symmetries $X=a_0 \partial_t + a_1 \partial_x + a_2
\partial_y$ or Lorentz transformations if we map the boundary $\I$
conformally to 3-dimensional Minkowski spacetime, i.e., in the Poincare
description~\eqref{poincare} of the conformal infinity.
Following the general prescription explained in Sec.~\ref{defs},
we first need to find the Noether charge corresponding to an infinitesimal
symmetry $X$. It is given by
\ben
({\bf Q}_X)_{\mu\nu} = \frac{1}{2} \nabla_\delta X_\rho
\epsilon^{\delta\rho}{}_{\mu\nu} \, .
\een
The symplectic 1-form potential is given by
\ben
\bt_{\mu\nu\sigma} = \left[
\frac{1}{2}(\nabla^\rho \delta g_\rho{}^\alpha - \nabla^\alpha \delta
g^\rho{}_\rho) - 2\delta A \nabla^\alpha A - 2\delta B \nabla^\alpha B
- i\overline{\chi} \gamma^\alpha \delta \chi + \dots
\right] \epsilon_{\alpha\mu\nu\sigma} \, ,
\een
where dots stand for other terms involving the fields $\psi_\mu, \chi,
A, B$ that vanish more rapidly at infinity and are not relevant for our analysis.
The conserved charge is then obtained by solving
\ben
\delta {\mathcal H}_X = \int \delta {\bf Q}_X - X \cdot \bt \, .
\een
where $\delta$ is an on-shell perturbation, and the integral is taken
over a cut at infinity.
To solve this equation for the Hamiltonian generators, we need to
know the asymptotic form of the on-shell variations
$\delta g_{\mu\nu}, \delta \psi_\mu, \delta A, \delta B, \delta \chi$.
This is determined by our boundary condition and by the field
equations, which impose additional constraints on the form of
the asymptotic expansion
of the metric. These constraints may be e.g. evaluated using the expansion
techniques of \cite{marolf}, see also \cite{Hertog2005,Amsel2006}:
The Gauss-Codacci relation
\ben
\tilde C_{\mu\alpha\nu\beta} \tilde n^\alpha \tilde n^\beta
= \pounds_{\tilde n} \tilde K_{\mu\nu} - \Omega^{-1} \tilde K_{\mu\nu}
+ \tilde K_{\mu\alpha} \tilde K^\alpha{}_\nu - \frac{1}{2}
\tilde h_\mu{}^\alpha \tilde h_\nu{}^\beta \tilde L_{\alpha\beta}
- \frac{1}{2} \tilde h_{\mu\nu} \tilde L^{\alpha\beta} \tilde n_\alpha
\tilde n_\beta
\een
relates the Weyl tensor of $\tilde g_{\mu\nu}$ to the extrinsic
curvature $\tilde K_{\mu\nu}$ of the $\Omega = const.$ surfaces
near infinity, and the matter stress tensor, where $\tilde L_{\mu\nu} =
T_{\mu\nu} - \frac{1}{3} g_{\mu\nu} T^\alpha{}_\alpha$. Combined with
the recursion relations~\eqref{recursion}, this gives
\ben
(h_{\mu\nu})_3 = - \frac{2}{3} (E_{\mu\nu})_0 +
\frac{2}{3}(K_{\mu\alpha}K^\alpha{}_\nu)_1
-\frac{1}{3} (\tilde h_\mu{}^\alpha \tilde h_\nu{}^\beta \tilde L_{\alpha\beta}
+\tilde h_{\mu\nu} \tilde L^{\alpha\beta} \tilde n_\alpha
\tilde n_\beta)_1 \, .
\een
Here, the subscripts indicate the order in
$\Omega$, and
\ben
E_{\mu\nu} = \Omega^{-1} \tilde C_{\mu\alpha\nu\beta} \tilde n^\alpha
\tilde n^\beta
\een
is the leading order part of the electric Weyl tensor.
Using the
asymptotic expansions of $A,B,\chi$ near infinity, and the
explicit form of the Lagrangian given above results in
\ben
(\tilde h_\mu{}^\alpha \tilde h_\nu{}^\beta \tilde L_{\alpha\beta}
+\tilde h_{\mu\nu} \tilde L^{\alpha\beta} \tilde n_\alpha
\tilde n_\beta)_1
= 2\tilde h_{\mu\nu} \left[ \frac{4}{3} (\alpha_A \beta_A + \alpha_B
  \beta_B) + \frac{i}{2} \bar \chi_- \Gamma \chi_+
\right] \, .
\een
Using this result, and eqs.~\eqref{recursion} and~\eqref{k1},
one can conclude that
the metric has the asymptotic expansion
\bena\label{asymptg}
\d s^2 &=& \frac{\d \Omega^2}{\Omega^2} -
\frac{1}{\Omega^2}
\left(1 - \left\{\frac{1}{2}\alpha_A^2 + \frac{1}{2} \alpha_B^2
                -\frac{1}{2} \right\}
          \Omega^2 \right) \, \d t^2 \non\\
&&    +\frac{1}{\Omega^2}\left(1 - \left\{ \frac{1}{2}\alpha_A^2 +
      \frac{1}{2} \alpha_B^2 - \frac{1}{2} \right\}
          \Omega^2 \right) \, \d s_{\mathbb
  S^2}^2  + \non\\
&& +\Omega \left\{ - \frac{2}{3} E_{\mu\nu} - 2\left[ \frac{4}{9}
(\alpha_A \beta_A + \alpha_B \beta_B) - \frac{1}{6} \bar \chi_- \Gamma \chi_+
\right] \tilde h_{\mu\nu} \right\} \d x^\mu \d x^\nu \non\\
&&\dots
\eena
Thus, the
asymptotic form of the on-shell perturbations for the fields may
be assumed to be
\bena\label{delg}
\delta g_{\mu\nu} &=& - \frac{2}{3} \Omega \delta E_{\mu\nu} + 2\Omega \delta \left[ \frac{4}{9}
(\alpha_A \beta_A + \alpha_B \beta_B) - \frac{i}{6} \bar \chi_- \Gamma \chi_+
\right] \tilde h_{\mu\nu} \nonumber \\
&& -\frac{1}{2}\delta (\alpha_A^2 + \alpha_B^2) \tilde h_{\mu\nu} +
\dots
\eena
as well as
\bena
\delta A &=& \Omega \delta \alpha_A + \Omega^2 \delta \beta_A +
\dots\\
\delta B &=& \Omega \delta \alpha_B + \Omega^2 \delta \beta_B +
\dots\\
\delta \chi &=& \Omega^{3/2} \delta \chi_+ - i\Omega^{3/2} \Gamma \delta
\chi_- + \dots \,
\eena
for the matter fields. The asymptotic form for the spin 3/2 field may
also be written out, but is not needed here.
If these expansions are used then one finds that
\bena
(A \nabla_\mu \delta A + B \nabla_\mu \delta B + \frac{i}{2} \chi
\gamma_\mu \delta \chi) \tilde n^\mu
&=& \frac{1}{2} \Omega\delta(\alpha_A^2 + \beta_B^2) + \Omega^2 \delta (\alpha_A \beta_A + \alpha_B
\beta_B) \\
&&+ \Omega^2 (\beta_A \delta \alpha_A + \beta_B \delta \alpha_A
+\frac{1}{2} \bar \chi_- \Gamma \delta \chi_+ - \frac{1}{2} \bar \chi_+ \Gamma \delta \chi_-).\non
\eena
The term in the last line is seen to vanish on account of our
boundary conditions on the fields $A,B,\chi$. Substituting this result
and~\eqref{delg} into the definitions of ${\bf Q}_X$, and of $X
\cdot \bt$, gives the result
\bena
(\delta {\bf Q}_X - X \cdot \bt)_{\mu\nu} &=&
\delta E_{\sigma\rho} X^\sigma \tilde n^\gamma \tilde
\epsilon^\rho{}_{\gamma\mu\nu} \\
&&+2\left[ \frac{1}{3}\delta(\alpha_A \beta_A + \alpha_B \beta_B) +
\frac{1}{2} \delta (\bar \chi_+ \Gamma \chi_-) \right] X^\sigma
\tilde n^\rho \tilde \epsilon_{\sigma\rho\mu\nu} \,  \nonumber
 \eena
at infinity. We now again use our boundary conditions on the scalar
and spinor fields. It follows that the expression for the conserved
charge associated with the symmetry $X$ is given by \bena\label{hX}
\H_X &=& -\int_C E_{\mu\nu} X^\nu u^\mu \d^2 S \\
&&+ \int_C \left\{
\frac{2}{3}
\beta_A [ {\mathcal F}'(\alpha_B) - \alpha_B {\mathcal
  F}''(\alpha_B) ] + [\alpha_B \F'''(\alpha_B) -
\F''(\alpha_B)] \bar \chi_+ \Gamma \chi_+  \right\} X^\mu u_\mu  \, \d^2 S \, ,
\non
\eena
where $\d^2 S$ is now the 2-dimensional integration element on
the cross section $C$ of $\I$ induced by the unphysical metric,
and $u^\mu$ is the timelike unit normal
(normalized with respect to the unphysical metric)
to $C$ within $\I$. The first term in $\H_X$ is the standard
gravitational contribution. The second
term arises from the backreaction of the matter fields onto the
metric and depends upon the function $\F$ specifying the boundary
conditions.
%The second term may be re-expressed noting that
%${\mathcal F}'(\alpha_B) = \alpha_A$, and that
%${\mathcal F}'(\alpha_B) - \alpha_B {\mathcal F}''(\alpha_B)$ is
%equal to the Legendre transformation $\widehat \alpha_A$ of $\alpha_A$,
%where the Legendre transformation $\widehat f(p)$
%of a function $f(q)$ is defined as usual by
%\ben
%\widehat f (p) = pq - f(q), \quad p = \frac{\d f(q)}{\d q} \, .
%\een
%Thus, we may write
This expression simplifies somewhat for the case of
$O(3,2)$-invariant boundary conditions $\F = \frac{1}{2}q
z^2$, in which case the Hamiltonian generators read
\ben
\label{HXq}
\H_X = -\int_C E_{\mu\nu} X^\nu u^\mu \d^2 S - q \int_C
\bar \chi_+ \Gamma \chi_+  \, X^\mu u_\mu  \, \d^2 S \, .
\een
  In this case, the matter term and the gravitational term
are separately conformally invariant.   It is interesting to note
that the scalars make no contribution to (\ref{HXq}).
For the special boundary conditions
found by Breitenlohner and Freedman~\cite{BF} (corresponding to
$q=0$ or $q=\infty$) the last term disappears as
well\footnote{To see this in the case
$q= \infty$ one should write the last term in terms of $\chi_-$.}.

\subsection{Positivity of the energy}

We now discuss the positivity properties of our expression for the
bosonic charges $\H_X$ given by eq.~\eqref{hX}, for timelike
asymptotic symmetries $X$, i.e., the positivity of the energy of the
theory. We first discuss the general boundary conditions on the
scalars preserving an $O(2,1) \times \mr^3$ bosonic asymptotic
symmetry group, specified by a general smooth function $\F$ as
in~\eqref{bndycond}. In that case, as follows from the discussion
below~\eqref{poincare}, a future directed timelike asymptotic
symmetry corresponds to a vector field $X$ given by
\ben\label{transl} X = a_0 \frac{\partial}{\partial t} + a_1
\frac{\partial}{\partial x} + a_2 \frac{\partial}{\partial y} \een
at $\I$, where $(a_0,a_1,a_2) \in \mr^3$ is a future directed,
timelike vector in 3-dimensional Minkowski spacetime and where
$(t,x,y)$ are the coordinates of $\I$ in the Poincare description of
the conformal boundary, see eq.~\eqref{poincare}. Such a vector
field can be written as \ben X = \overline \varepsilon \gamma^\mu
\varepsilon \een where $\varepsilon$ is a complex linear combination
of the 2 special linearly independent asymptotically Killing
spinors\footnote{Unlike in the rest of the paper, spinors in this
subsection are taken to be commuting. The reason is that we
want the spinor charge $Q_X$ defined below to be a real number. The
spinors considered here are not dynamical fields.} described below
eq.~\eqref{poincare}. We will now show that \ben \H_X \ge 0 \een for
such vector fields, for any solution of the field equations with the
stated boundary conditions for which the spinor fields all vanish
\ben \label{nospin} \psi_\mu = 0 = \chi \, .
 \een
The condition (\ref{nospin}) guarantees that $\H_X$ is real, as
opposed to being merely an even element of the exterior (Grassmann) 
algebra. 

The Lagrangian~\eqref{lagr} then reduces to its bosonic part, given
by \ben \LL_{bos} = \left[ \frac{1}{2} R - G_{IJ}(\phi) \nabla^\mu
\phi^I
  \nabla_\mu \phi^J - V(\phi)
\right] \beps \, . \een Here, the scalar fields $A,B$ have been
combined into a 2-component field $\phi=(A,B)$ taking values in a
2-dimensional target space with metric and potential given by \ben
G_{IJ}(\phi) \d \phi^I \d \phi^J = \frac{|\d
\phi|^2}{(1-|\phi|^2)^2}, \quad V(\phi) = -
\frac{3-|\phi|^2}{1-|\phi|^2} \, . \een The potential arises from
the pre-potential $P(\phi)=(1-|\phi|^2)^{-1/2}$ in the sense that
there holds the equation \ben\label{prep} V = -3P^3 + G^{IJ}
\partial_I P \partial_J P \, , \een which is closely related to the
fact that $\LL_{bos}$ comes from a supersymmetric theory.
Following~\cite{Townsend1981}, generalizing in turn arguments
of~\cite{Boucher1979,Gibbons1983,Gibbons1984,Witten1980,Nester1982}, 
one can show that the
existence of such a prepotential entails the existence of a positive
definite quantity $Q_X \ge 0$, which will turn out to be equal to
$\H_X$, thus establishing the desired positivity property. The
quantity $Q_X$ is defined as follows. One first defines an analog of
the Nester~\cite{Nester1982} 2-form, given by \ben\label{nester}
F^{\alpha\beta} = \bar \varepsilon \gamma^{[\alpha} \gamma^\beta
\gamma^{\rho]} \left( \nabla_\rho + \frac{1}{2} P(\phi) \gamma_\rho
\right) \varepsilon \, . \een This quantity will turn out to be
closely related to the fermionic supercharges of the theory, but for
the moment we do not need this fact. The quantity $Q_X$ is now
defined as
 \ben Q_X = \int_C {}^* {\bf F} \, .
  \een
 Let us choose a
spacelike 3-surface intersecting the cut $C$ of $\I$ transversally,
with no interior boundary (we assume that such a surface can be
chosen). Then, applying Stoke's theorem, we have
 \ben Q_X =
 \int_\Sigma \d (^* {\bf F}) = \int_\Sigma (\nabla^\mu F_{\mu\nu})
u^\nu \, dS \, ,
 \een where $dS$ is the natural integration element
on $\Sigma$, and where $u^\mu$ is the normal, normalized to -1 here
with respect to the physical metric. Using~\eqref{prep} and
Einstein's equation, one can calculate~\cite{Townsend1981}, that
\ben\label{divF} (\nabla^\mu F_{\mu\nu}) u^\nu = -2 (\widehat
\nabla_i \varepsilon)^\dagger \gamma^i \gamma^j \widehat \nabla_j
\varepsilon + 2(\widehat \nabla_i \varepsilon)^\dagger \widehat
\nabla^i \varepsilon + G_{IJ} \lambda^{I
  \, \dagger} \lambda^J \, ,
\een
where $i,j$ are indices associated with the tangent space of $\Sigma$,
where $\widehat \nabla_\mu = \nabla_\mu + \frac{1}{2} P(\phi)
\gamma_\mu$, and where
\ben
\lambda^I = \frac{1}{\sqrt{2}} ( \gamma^\mu \nabla_\mu \phi^I + 2
G^{IJ} \partial_J P) \varepsilon \, .
\een
The last 2 terms in eq.~\eqref{divF} are manifestly positive,
while the first term can be set to 0 by imposing the ``Witten condition''
$\gamma^i \widehat \nabla_i \varepsilon = 0$ on $\Sigma$ (essentially the
spatial Dirac equation with suitable boundary conditions).
As one can show using functional analytic techniques~\cite{Hertog2005}
(see also~\cite{Amsel2006b}) a
global, smooth solution to this condition with the desired boundary conditions
indeed exists under our assumed boundary conditions. Thus,
we have shown that $Q_X \ge 0$.

We next need to relate $Q_X$ to $\H_X$. For this, we expand the
metric and scalar field as above in~\eqref{asymptg} and~\eqref{ABas}
near $\I$, and we additionally expand $\varepsilon$ near $\I$ using
the fact that it satisfies the Witten condition, see
\cite{Hertog2005,Amsel2006} and~\cite{Amsel2006b} for similar
calculations in an almost identical theory with only one scalar
field. Using our boundary conditions on the scalar
fields~\eqref{bndycond}, as well as an expansion of $P$ near
$\phi=0$ one finds\footnote{  Here it is important that $P$ is a
``good'' superpotential of the type called ``$P_-$''
in~\cite{Amsel2006b}, see that paper for a detailed explanation.  }
\ben (^* {\bf F})_{\mu\nu} = \tilde \epsilon_{\mu\nu\alpha\beta}
\tilde n^\beta \left[ E^{\alpha \rho} X_\rho - \frac{2}{3} (\alpha_A
\beta_A + \alpha_B \beta_B) X^\alpha \right] \, \een at $\I$.
Integrating this expression over a cut $C$ at $\I$, again using our
boundary conditions on the scalar fields, and comparing the result
to our expression~\eqref{hX} for $\H_X$ with the fermionic fields
$\chi$ set to 0, one sees that $Q_X = \H_X$. Thus, we have shown
that $\H_X \ge 0$ for timelike asymptotic symmetries and boundary
conditions preserving an $O(2,1) \times \mr^3$ group of asymptotic
symmetries.

The choice $\F(z) = \frac{1}{2} q z^2$ is a special case which,
as explained above, corresponds to an asymptotic symmetry group
isomorphic to $O(3,2)$. In that case, we have a larger class of
timelike asymptotic Killing fields, which can be written as $X^\mu =
\overline \eta \gamma^\mu \eta$, where $\eta$ is a complex linear
combination of the 4 real linearly independent asymptotic Killing
spinors rather than only a linear combination of the special 2 real
linear independent spinors $\varepsilon$ described
below~\eqref{poincare}. Each such $\H_X$ is positive by the argument
above. 

\subsection{Fermionic charges}

Since we have identified a set of boundary conditions preserving
${\mathcal N}=1$  supersymmetry and enforcing
vanishing symplectic flux through $\I$, we know by the
analysis of Sec.~\ref{defs} that
corresponding conserved Hamiltonian generators associated with
a supersymmetry variation exist. We now explicitly determine
their form. As in the previous subsection, it is
instructive to treat separately the special boundary conditions
defined by $\F(z) = \frac{1}{2}q z^2$, preserving a group $O(3,2)$
of asymptotic bosonic symmetries, and the general boundary conditions
defined by a general $\F(z)$, preserving only a subgroup $O(2,1) \times
\mr^3$ of asymptotic bosonic symmetries.

In the general case, asymptotic fermionic symmetries are associated
with asymptotic Killing spinors $\varepsilon$ which are linear
combination of the 2 special linearly independent spinors described
below eq.~\eqref{poincare}. For such spinors, the asymptotic Killing
field $X^\mu =\bar \varepsilon \gamma^\mu \varepsilon$ is a
translation on $\I$, i.e., of the form~\eqref{transl}, corresponding
to generators in the translation subgroup of $O(2,1) \times \mr^3$.
To determine the fermionic Hamiltonian generators $\H_\varepsilon$
following the general procedure outlined in Sec.~\ref{defs}, we must
determine the Noether charge, ${\bf Q}_\varepsilon$ [see
eq.~\eqref{noether}] and the 2-form ${\bf A}_\varepsilon$ [see
eq.~\eqref{Adef}] for extended supergravity, and then determine
$\H_\epsilon$ via eq.~\eqref{Hdef}. From the general form of the
expressions that are involved in these quantities, one can infer
that the integrand ${\bf h}_\varepsilon$ of the Hamiltonian
generator $\H_{\varepsilon} = \int_C {\bf h}_\varepsilon$ must be a
2-form satisfying the following general properties: (1) It must be
locally constructed out of $\varepsilon$, the metric $g_{\mu\nu}$,
the Levi-Civita tensor $\epsilon_{\mu\nu\sigma\rho}$, the
gamma-matrices $\gamma_\mu$, the spin 3/2-field  $\psi_\mu$, and the
matter fields $A, B, \chi$, but not the covariant derivatives of
these fields. (2) It must be linear in $\varepsilon$. (3) It must be
manifestly finite at $\I$. These restrictions only leave a
relatively small number of possible terms, since for example each
factor of $A$ or $B$ drops off as $\Omega$ near infinity, so that
terms involving higher powers of $A$ or $B$ will automatically
vanish at $\I$. In fact, it can be seen in this way that possible
terms fulfilling (1), (2) and (3) can at most be linear in
$\psi_\mu, \chi, A, B$, and cannot depend upon $A,B,\chi$ if they
depend upon $\psi_\mu$. One possible term fulfilling these criteria
is the 2-form $^* {\bf N}_\varepsilon$ entering the definition of
the Hamiltonian generator in minimal supergravity (see
eq.~\eqref{N}). Let us define \ben {\mathcal G}_\varepsilon = \int_C
{}^* {\bf N}_\varepsilon \, . \een We will now show that, in fact,
\ben\label{heta} \H_\varepsilon = {\mathcal G}_\varepsilon \een
i.e., the formula for the Hamiltonian generators for the fermionic
charges in extended supergravity is the same as in minimal
supergravity.

We could prove this directly by going stubbornly through the
definitions of sec.~\ref{defs}, but this is somewhat tedious
because the Lagrangian~\eqref{lagr} and the corresponding
supersymmetry transformations are relatively complicated. We will instead
use an indirect argument proving that $\H_\varepsilon = {\mathcal
  G}_\varepsilon$. This argument is based on
our knowledge of the bosonic charges $\H_X$ associated with an
asymptotic symmetry vector field $X$, see eq.~\eqref{hX}. The point
is that, since the generators $\H_\varepsilon$ and $\H_X$ are
defined as Hamiltonian generators, they form a representation of the
superalgebra of asymptotic symmetries (isomorphic to the super
Poincar\'e algebra in 3 dimensions) on the covariant phase space
under the Poisson bracket (with respect to the symplectic form
$\sigma$), that is, \ben\label{alg} \{\H_{\varepsilon_1},
\H_{\varepsilon_2}\} = \H_X \, , \quad X^\mu = \bar \varepsilon_1
\gamma^\mu \varepsilon_2 \, . \een Thus, the fermionic charges are
related to the bosonic ones, whose form is already known from the
previous section. That relation will enable us to derive the desired
expression for the fermionic generators in extended
supergravity~\eqref{heta}. We will show that \ben\label{qcom}
\{\H_{\varepsilon_1}, {\mathcal G}_{\varepsilon_2}\} = \int_C
\b_{\varepsilon_1} \, {}^* {\bf N}_{\varepsilon_2} = \H_X \, . \een
In view of eq.~\eqref{alg}, this means that $\H_\varepsilon$ can
differ from ${\mathcal G}_\varepsilon$ at most by a term ${\mathcal
D}_\varepsilon$ with vanishing supersymmetry variation. We now argue
that there is no such term. Indeed, by properties (1), (2), and (3),
the 2-form integrand of ${\mathcal D}_\varepsilon$ would have to be
a local functional of the dynamical fields (but not their
derivatives) that is linear in $\varepsilon$, and at most linear in
$\psi_\mu, A, B, \chi$. The possible terms can be written out
explicitly, and it can be seen using
eqs.~\eqref{bsusy1}-\eqref{bsusy2} that there does not exist a term
which has a vanishing supersymmetry variation (up to an exact
2-form). Thus, if we can prove eq.~\eqref{qcom}, then
eq.~\eqref{heta} follows.

The supersymmetry variation of the 2-form
$^* {\bf N}_{\varepsilon_2}$ on $\I$ in expression~\eqref{qcom}
can be computed using the definition of
the supersymmetry transformation on the spin-3/2 field, giving at $\I$
\bena\label{bN}
(\b_{\varepsilon_1} \, {}^* {\bf N}_{\varepsilon_2})_{\mu\nu}  &=&
\epsilon^{\alpha\beta}{}_{\mu\nu} \, \bar \varepsilon_1 \gamma_{[\alpha}
\gamma_\beta \gamma_{\rho]}
\left( \nabla^\rho + \frac{1}{2} P(A,B) \, \gamma^\rho \right) \varepsilon_2
\non\\
&&+
%\frac{1}{2}
(A \nabla_{[\mu} B - B \nabla_{[\mu} A) X_{\nu]} \, ,
\eena
where $P$ is the prepotential given above in eq.~\eqref{prep}. An
important thing to note is that the first term on the right side is
essentially the dual of the Nester 2-form given above in
eq.~\eqref{nester}. Thus, by essentially the same argument as given
there, the integral over a cut $C$ of the first term will give us
$\H_X$. We only need to take care that the fields $\chi, \psi_\mu$ are now
not assumed to vanish, and consequently might appear e.g. in the
asymptotic form of the metric~\eqref{asymptg}.
Taking into account this difference,
we obtain for the first term in~\eqref{bN} the expression
\ben\label{22}
\text{(First term)} = \tilde \epsilon_{\mu\nu}{}^{\alpha\beta} \tilde n_\beta
\left[ E_{\alpha\rho} X^\rho - \frac{2}{3}
(\alpha_A \beta_A + \alpha_B \beta_B - \frac{3}{2}
\bar \chi_+ \Gamma \chi_-) X^\alpha \right] \, .
\een
On the other hand, the second term in~\eqref{bN} can be expressed
as
\ben\label{33}
\text{(Second term)} =
%\frac{1}{2}
\tilde X_{[\nu} \tilde \nabla_{\mu]} \{ 2\F(\alpha_B) - \alpha_B
\F'(\alpha_B) \} \een using the boundary
conditions~\eqref{bndycond}, where $\tilde X_\mu = \tilde g_{\mu\nu}
X^\nu$ is given by \ben\label{transl1} \tilde X = a_0 \d t + a_1 \d
x + a_2 \d y \een in view of eq.~\eqref{transl}. We now integrate
eq.~\eqref{bN} over a cut $C$ of $\I$, and compare the result with
our expression for the Hamiltonian generator~\eqref{hX}
using~\eqref{22} and~\eqref{33}.  We find \ben\label{mainr}
\{\H_{\varepsilon_1}, {\mathcal G}_{\varepsilon_2}\} = \H_X + \int_C
\d [ 2\F(\alpha_B) - \alpha_B \F'(\alpha_B)] \wedge \tilde X \, .
 \een
Integrating the last term by parts and computing $\d \tilde X = 0$
from~\eqref{transl1} establishes the desired relation~\eqref{qcom}.

\medskip

We next consider the case of the special $O(3,2)$-invariant boundary
conditions corresponding to the special choice $\F(z) = \frac{1}{2}
q z^2$. In that case, our supersymmetry transformations $s_\eta$ are
parametrized by arbitrary asymptotic Killing spinors $\eta$, and not
the special 2-dimensional subspace of asymptotic Killing spinors
$\epsilon$ giving rise to translational asymptotic Killing fields
$X$, see~\eqref{transl}. In that case, eq.~\eqref{mainr} still holds
(with $\varepsilon_i$ replaced by the more general spinors
$\eta_i$), but it is no longer true that $\d \tilde X = 0$ on $\I$,
because $X$ need no longer be a translation on $\I$. However, we now
have
 \ben
  2\F(\alpha_B) - \alpha_B \F'(\alpha_B) = 0 \, .
  \een
Thus, the second term in eq.~\eqref{mainr} still vanishes, 
demonstrating that the desired relation~\eqref{qcom} also holds
(with $\epsilon_i$ replaced by $\eta_i$ in that equation) for the
larger set of supersymmetry generators associated with $\F(z) =
\frac{1}{2} q z^2$.

\section{Summary and conclusions}

 In this paper we have investigated classical field theories with
local symmetries, and in particular theories with fermionic local symmetries, i.e., supergravity theories. Let us summarize the main results of this paper:

\begin{itemize}
\item
In section~2, we have gave a general definition of Hamiltonian
generators conjugate to a local gauge symmetry
in classical field theories on a manifold, within a covariant phase
space framework.
The local symmetries were assumed to be either of bosonic type
(such as local Yang-Mills type gauge symmetries or diffeomorphisms)
or of fermionic type (such as supersymmetry transformations).

For theories requiring the specification a set of boundary and/or
asymptotic conditions, the existence of the generators is not
guaranteed. We derived a simple general criterion  stating when such
charges exist and are conserved. The criterion is that the
symplectic flux through any real or conformal boundary should
vanish, and that the boundary conditions should be invariant under
the symmetry. In the case of a bosonic symmetry, our method reduces
to that of Wald et al.~\cite{Iyer1994,Iyer1995,Zoupas1998}
\item
We then applied our formalism to two example theories: Minimal ($\N=1$)
supergravity with a negative cosmological constant, and a consistent
truncation of extended $\N=4$ supergravity with a complex scalar
field of mass $m^2=-2$, a Majorana fermion field, and a cosmological
constant $\Lambda = -3$. For minimal supergravity, we chose boundary
conditions
 such that the metric approaches that of
AdS-spacetime for large distances, and in particular such that the boundary metric is held fixed.   We then found boundary conditions for the
spin-3/2 field such that our criteria for the existence of
supercharges were met. In particular, those boundary conditions are
invariant under the supergroup of the asymptotic symmetry group
$O(3,2)$. An explicit formula for the supercharge was
also derived, and it agreed with the formula previously found by
Henneaux and Teitelboim~\cite{Henneaux1985} by a somewhat different
method.

\item
For extended $\N=4$ supergravity theory, one also has to
choose boundary conditions for the matter fields. We found a
1-parameter family of boundary conditions perserving $\N=1$
supersymmetry, satisfying our criteria for the existence of the
corresponding $\N=1$ supercharges. These boundary conditions
are, in particular invariant under the $\N=1$ supergroup of $O(3,2)$.
For particular values of the parameter, our boundary conditions
reduce to those found by Breitenlohner and Freedman~\cite{BF}.
We also found a more general class of boundary conditions of
non-linear type that are invariant under a subgroup $O(2,1) \times \mr^3
\subset O(3,2)$, and which are also invariant under a restricted set
of supersymmetry transformations generating
 Poincare transformations of the
asymptotic boundary $\mr^3$ of spacetime (corresponding to the
description of AdS-spacetime as
conformal to a half space of 4 dimensional
Minkowski space). These boundary conditions are characterized by
an arbitrary function, $\F(z)$, in one variable. They could be written as
 a ``supersymmetric Lagrange submanifold condition''
 \ben
 \Phi_+ = \frac{\delta}{\delta \Phi_-} \int_\I \F(\Phi_-) \, \d^3 S \d^2 \theta
\, ,
 \een
where $\Phi_\pm$ are boundary superfields on the conformal
boundary $\I$ of spacetime, defined from the asymptotic
tails of the scalar and spin-1/2 fields in the theory. The boundary
conditions of~\cite{BF} correspond to the particular choice $\F(z)=0$.
Asymptotic supercharges were shown to
exist (for general $\F$) that generate the restricted class of supersymmetry
transformations. In the context of the AdS/CFT correspondence, the
boundary conditions
 associated with  a general $\F$ correspond to a
modified boundary action by general
arguments~\cite{Berkooz:2002ug,Sever:2002fk,Witten:2001ua},
\ben
S_{bndy} \to S_{bndy} + \int_\I \F({\mathcal O}) \, \d^3 S \d^2 \theta \,
,
\een
with $\mathcal O$ the dual boundary superfield in the CFT.
This modification will in general break superconformal invariance,
but preserves supersymmetries corresponding to the 3-dimensional
Poincare group. Our boundary conditions thus give rise to interesting
QFT-duals.

\item
 Finally, we gave formulae for the bosonic and fermionic charges in
extended $\N=4$ supergravity, and we proved that
the energy is always positive when the fermionic fields are set to 0.
Our formulae for the fermionic
charges agree with those in minimal $\N=1$ supergravity,
though the explicit expressions for the bosonic charges generally contain an additional term associated with the matter fields.  This additional term depends upon
the function $\F$ characterizing the boundary conditions.

\end{itemize}

Although we have focused on particular supergravity theories, the
general pattern of results is clear.  In the ${\cal N}=4$ theory,
the freedom to choose a variety of boundary conditions preserving
${\cal N}=1$ supersymmetry stems from the presence of appropriate
matter chiral superfields.  One therefore expects a similar set of
allowed ${\cal N}=1$ boundary conditions in any $d=4$ theory
containing chiral superfields with masses in the usual window
\cite{BF} above the Breitenlohner-Freedman bound. This result
is also suggested by the analysis of \cite{Dmitry}.  It would be
interesting to check this, as well as the details of generalizations
to other dimensions. In particular, for $d \neq 4$ it remains to
determine the form of any associated boundary superfields.  We note
that a generalization to $d=5$ would address a number of interesting
examples already well-studied in the context of AdS/CFT
\cite{PS,PW,LM}.

Let us return, however, to $ {\cal N}=4$ supergravity in $d=4$.  Even here, one expects to obtain an additional class of ${\cal N} =1$ theories by deforming the boundary conditions for the vector super-multiplet.  Such a possibility was indicated in the original work of Breitenlohner and Freedman \cite{BF}, and in the context of AdS/CFT would amount to a supersymmetric generalization of \cite{WittenV, LP1,LP2,Petkou,Yee,MR}.  We have not considered this option here, but again it would be of interest to work out the full formalism and to obtain a useful form of any associated vector ``boundary superfields."

Finally, we have considered only boundary conditions for the graviton super-multiplet which impose the usual boundary condition on the graviton itself: namely, that the induced metric on the boundary is fixed.  However, for $d=4$
one may also impose boundary
conditions \cite{IW} which have slower fall-off near $\I$.
In this case, AdS/CFT arguments \cite{LP1,LP2,MR} suggest that such boundary
conditions are related to ``dual'' boundary theories which {\it also} contain gravity and which are defined on
spacetimes with compact Cauchy surfaces.  As a result, one expects that
with such slower fall-off boundary conditions, all asymptotic
symmetries will be gauge and lead only to trivial conserved charges.  It would, however, be interesting to check that supersymmetry can be preserved in such contexts, and to check that indeed the associated variations $\delta \H_\eta, \delta \H_X$ vanish on-shell.

\section*{Acknowledgements}
The authors would like to thank Rob Leigh for useful discussions.
D.M. was supported in part by NSF
grant PHY0354978, by funds from the University of
California. S.H. would like to thank the physics department
at UC Santa Barbara for their kind hospitality.

%%%%%%%%%%%%%%%%%%%%%%%%%%
%%%%%%%%%%%%%%%%%%%%%%%%%%

\appendix

\section{General properties of $\H_\eta$}

In this appendix, we summarize some important, model independent
general properties of the generators $\H_\eta$ for any
(bosonic or fermionionic) symmetry $\eta$.

\subsection{Ambiguity of $\H_\eta$:}
Two potential ambiguities arise in the
construction of $\H_\eta$. First, a potential ambiguity arises
from the fact that we may change the Lagrangian $\LL$ to
$\LL + \d {\bf \mu}$, with
${\bf \mu}$ a $(d-1)$-form that is locally constructed out of the fields,
without changing the field equations. However, it is easy to see
that such a change will not affect the symplectic form
$\bomega$, and hence will not change the definition $\H_\eta$.
Another potential ambiguity arises because
$\bt$ is only unique up to the addition of an exact form,
\ben
\bt(\Phi, \delta \Phi) \to \bt(\Phi, \delta \Phi) + \d \,
{\bf Z}(\Phi, \delta \Phi) \, .
\een
The addition of such a term will change the definition of $\H_\eta$---assuming
it exists---by
\ben
\H_\eta \to \H_\eta + \int_{\partial \Sigma} {\bf Z}(\Phi, \b_\eta \Phi) \, .
\een
This type of ambiguity will typically be eliminated by the given
choice of boundary conditions, which ensure
 that  the only  ${\bf Z}$s which gives rise to finite modification of the
generator $\H_\eta$ simply shift  $\H_\eta$ by constants, if they exist at all.

\subsection{Gauge invariance of $\H_\eta$}
Note that $\Q_\eta$ and $\A_\eta$ in
the definition of $\delta \H_\eta$ are by construction ``potentials''
for gauge invariant quantities, and therefore
need not be gauge invariant themselves.
Consequently $\H_\eta$ itself is in danger of not being gauge invariant.
We claim, however, that $\delta \H_\eta$, and hence $\H_\eta$ itself,
is always invariant under gauge transformations that can be continuously connected
to the identity (i.e., except possibly for ``large gauge transformations'').
As an example, consider the case when one of the matter fields is
a non-abelian gauge field $A_\mu$. If we assume that the
Lagrangian is invariant under infinitesimal gauge transformations
$\delta A_\mu = \nabla_\mu \Lambda + [A_\mu, \Lambda]$,
then it follows (by going through the definitions),
that $\Q_\eta$ is invariant under such a transformation
up to the addition of a closed from ${\bf G}_\Lambda$. However, again by
the fundamental lemma, it follows that ${\bf G}_\Lambda$ must be exact, and
hence cannot contribute to the integral defining $\delta \H_\xi$. A similar
argument can be made for the second term contributing to this integral.
Hence, $\delta \H_\eta$, and therefore $\H_\eta$ itself, is independent under
infinitesimal gauge transformations. But this implies that it is
independent under any finite gauge transformation that is connected to the
identity by a differentiable path.

\subsection{Off-shell formula for $\H_\eta$}

 It is often useful to have an off-shell formula for $\H_\eta$, for example, for the computation of Poisson brackets with interesting gauge dependent quantities.  To construct such a formula, we
now compute $\delta \H_\eta(\Phi)$ for any solution $\Phi$,
but arbitrary off shell variations $\delta \Phi$.
Note that, in this situation, our derivation of the
variation of the generator $\H_\eta$ only implies that
\ben\label{5}
\delta \H_\eta = \int_\Sigma \delta \J_\eta + \int_{\partial \Sigma} \A_\eta
\, .
\een
On shell, we have $\J_\eta = \d \, \Q_\eta$, so
we also have $\delta \J_\eta = \d \, \delta \Q_\eta$ for
on shell variations $\delta \Phi$, and the first
integral reduces to a surface integral~$\int \delta \Q_\eta$.
However, this is not true for off-shell variations,
and the off-shell variation
of $\H_\eta$ consequently
differs  from~\eqref{5} by
a term that is not a boundary integral. To characterize this term,
it is necessary to have a suitable
off-shell formula for $\Q_\eta$.
Since $\J_\eta = \d \Q_\eta$ only holds on shell, we have in general
\ben\label{4}
(\J_\eta - \d  \Q_\eta)_{\nu_1 \dots \nu_{d-1}} = \sum_{I=0}^N
\lambda^{I \, (\mu_1 \dots \mu_I)}{}_{A [\nu_1 \dots \nu_{d-1}]} (\Phi)
\nabla_{(\mu_1} \cdots \nabla_{\mu_I)} \eta^A \, \, ,
\een
where $\lambda^I(\Phi)=0$ vanishes if $\Phi$ is on shell. We will now
show that, if the variation $\b_\eta \Phi$ contains at most
$M$ derivatives of $\eta^A$, then the Noether charge can be modified
$\Q_\eta \to \Q_\eta + \btau_\eta$ by a $(d-2)$-from $\btau$ vanishing
on shell, such that the right side of eq.~(\ref{4}) contains at
most $M-1$ derivatives of $\eta^A$. The proof of this statement is similar 
in nature to arguments given in~\cite{Wald1988}: Let the maximum number $N$ of
derivatives in eq.~(\ref{4}) be such that $N \ge M$. Since
\ben
\d (\J_\eta - \d  \Q_\eta) = \E \, \b_\eta \Phi \, ,
\een
and since the right side of that equation by assumption
contains at most $M$ derivatives of $\eta^A$, it follows
that
\ben
0 =
\lambda^{(\mu_1 \dots \mu_N}{}_{A \, [\nu_1 \dots \nu_{d-1}}
\delta^{\sigma)}{}_{\rho]}
\nabla_{(\mu_1} \cdots \nabla_{\mu_N} \nabla_{\sigma)}
\eta^A
\een
for all $\eta^A$. Consequently, we have
\ben
0 =
\lambda^{(\mu_1 \dots \mu_N}{}_{A \, [\nu_1 \dots \nu_{d-1}}
\delta^{\sigma)}{}_{\rho]} \, .
\een
Defining now
\ben
\tau_{\nu_1 \dots \nu_{d-2}} = \frac{N}{N+1}
\lambda^{\sigma \mu_1 \dots \mu_{N-1}}{}_{A \, \sigma
[\nu_1 \dots \nu_{d-2}]}
\nabla_{(\mu_1} \cdots \nabla_{\mu_{N-1})}
\eta^A \, ,
\een
it may be checked that $\J_\eta - \d \Q_\eta - \d \btau_\eta$ contains
at most $N-1$ derivatives of $\eta^A$, and clearly $\btau(\Phi) = 0$
on shell. Therefore, the highest derivative term on the right side
of eq.~(\ref{4}) has been removed by an off-shell redefinition of $\Q_\eta
\to \Q_\eta - \btau_\eta$
which does not affect the form of $\Q_\eta$ on shell.
We can continue this process
inductively to remove the next-highest derivative term on the right
side of~(\ref{4}), as long as the number of derivatives is not smaller
than $M$. Therefore, by a suitable off-shell redefinition of $\Q_\eta$,
it can always be achieved that
\ben
\J_\eta - \d  \Q_\eta = \sum_{I=0}^{M-1} \C_A{}^{\mu_1 \dots \mu_I}(\Phi)
\nabla_{(\mu_1} \cdots \nabla_{\mu_I)} \eta^A \, ,
\een
where $\C^A(\Phi)$ vanishes on shell, and where $M$ is the maximum number
of derivatives of $\eta^A$ appearing in the symmetry variation $\b_\eta \Phi$.
Therefore, by eq.~\eqref{5}, the total off-shell Hamiltonian variation
is given by
\ben
\delta \H_\eta =
\sum_{I=0}^{M-1} \int_\Sigma \delta \Bigg[
\C_A{}^{\mu_1 \dots \mu_I}
\nabla_{(\mu_1} \cdots \nabla_{\mu_I)} \eta^A \Bigg] + \int_{\partial \Sigma}
\delta \Q_\eta-\A_\eta \, .
\een
The forms $\C_A{}^{\mu_1\dots \mu_I}$ are the ``constraints'' associated with the
fermionic symmetries, and vanish when the equations of motion hold. Thus,
we see that the off shell variation of $\H_\eta$ is not given by
a boundary integral, but contains also a ``bulk'' integral
containing the variation of the constraints of the theory associated
with the fermionic symmetries.
In particular, if the number of derivatives appearing
in the SUSY transformations is $M=1$, as happens e.g. when the Lagrangian
contains no more than second derivatives of the fields (the
case in nearly all applications, and in particular in the
supergravities studied in this paper),
then
\ben
\J_\eta - \d  \Q_\eta = \C_A \eta^A \, ,
\een
and we get a correspondingly simpler form for the
off shell variation of $\H_\eta$,
\ben\label{offshell}
\delta \H_\eta = \int_\Sigma \delta(
\C_A \eta^A)  + \int_{\partial \Sigma}
\delta \Q_\eta-\A_\eta \, .
\een

%---------------------------------------------------------------------------

\end{document}